\documentstyle[preprint,eqsecnum,aps]{revtex} 
\begin{document}
\draft
\title{On the multipole expansion of the gravitational field}
\author{Luc BLANCHET}
\address{D\'epartement d'Astrophysique Relativiste et de Cosmologie,\\
Centre National de la Recherche Scientifique (UPR 176),\\
Observatoire de Paris, 92195 Meudon Cedex, France}
\date{\today}
\maketitle

\begin{abstract}
This paper constructs the multipole expansion (in general relativity) of the gravitational field generated by a slowly-moving isolated source. We introduce some definitions for the source multipole moments, valid to all orders in a post-Newtonian expansion, and depending in a well-defined way on the total stress-energy pseudo-tensor of the material and gravitational fields. Previously obtained expressions of the source multipole moments are recovered in the appropriate limits. The source moments parametrize the linearized approximation of the gravitational field exterior to the source, as computed by means of a specific post-Minkowskian algorithm defined in a previous work. Since the radiative multipole moments parametrizing 
the radiation field far from the source can be obtained as non-linear functionals of the source moments, the present paper permits relating the radiation field far from a slowly-moving source to the stress-energy pseudo-tensor of the source. This should be useful when comparing to theory the future observations of gravitational radiation by the LIGO and VIRGO experiments.
\end{abstract}

\section{Introduction} \label{sec:1}

The multipole expansion is one of the most useful tools of theoretical
Physics.  Extensively used in the past for dealing with wave emission
and/or propagation problems in electromagnetism, the multipole
expansion has also much contributed to gravitational radiation theory,
starting with Einstein's pioneering work \cite{E18} (see also \cite{LL})
showing
that gravitational radiation from a localized slowly moving source is
dominantly quadrupolar, unlike electromagnetic radiation which is
dipolar in general.

The multipole expansion of the field generated by an isolated
gravitating source can be said to be fully understood at the {\it
linearized} approximation of general relativity.  Some early studies
extended the Einstein quadrupole formula for the energy to include
higher multipole contributions \cite{Papa62,Papa71}, and obtained the
corresponding formulas for linear momentum
\cite{Papa62,Papa71,Bek73,Press77} and angular momentum
\cite{Pe64,CB69}.  Then a number of authors \cite{SB58,S61,Pi64,Th80}
investigated the general structure of the {\it infinite} multipole
expansion of the linearized field outside the source, and computed the
associated fluxes of energy and momenta.  It emerged from these works
that the multipole expansion is entirely characterized by two and only
two sets of multipole moments, which are the analogues of
the electric and magnetic moments of electromagnetism and are
referred in the gravitational case to as the mass and current moments.
Notably the full expression of the linearized vacuum metric as
parametrized by symmetric and trace-free (STF) mass and current
multipole moments was given by Thorne \cite{Th80} (we shall refer
below to this expression as ``canonical'').  The use of STF harmonics is
very convenient when doing computations of gravitational radiation
\cite{Th80}.

Knowing the structure of the multipole expansion is
not sufficient by itself, and one must relate the multipole moments to
the material content of the source.  In linearized gravity this
relationship was achieved in a second series of papers
\cite{M62,CM71,CMM77,DI91b}, in which the multipole moments of the source were
obtained as some explicit integrals extending over the stress-energy
distribution of the matter fields, independently of any assumption
concerning the source (it be fast moving for instance).  The last of
these papers, by Damour and Iyer \cite{DI91b}, put the finishing touches
by expressing the multipole expansion in STF form, and by correcting
some errors in the previous paper \cite{CMM77}.

The problem is more difficult, but more interesting, in the full
nonlinear theory where the (source) multipole moments 
mix with each other causing
non-linear effects in the radiation field, for instance the so-called
``tail'' effect --~scattering of the quadrupolar waves off the spacetime
generated by the mass monopole of the source.  Several approaches have
attempted to define a multipole expansion consistent with the nonlinear
field equations. In particular, when the
source is stationary (existence of a time-like Killing
vector), the multipole moments in the far field are determined by the
convergent expansion of the metric at spatial infinity ($r \to +\infty$,
$t=$~const). Several equivalent definitions of the far-field 
moments are named
after Geroch and Hansen \cite{G70,H74}, Thorne \cite{Th80} and Simon and
Beig \cite{SB83}.  In the non-stationary case, the expansion at future
null infinity ($r \to +\infty$, $u=$~const where $u$ is a null coordinate)
was constructed by Bondi {\it et al} \cite{BBM62} and Sachs \cite{S62}.
By decomposing the so-called Bondi news function \cite{BBM62} in
spherical harmonics (or in STF harmonics), one obtains the multipole
moments which are actually ``measured'' at
infinity (radiative multipole moments). Another attempt at defining the multipole expansion for non-stationary sources is that of Bonnor and collaborators \cite{Bo59,BoR61,BoR66,HR69}, who combined the multipole expansion with a weak-field or post-Minkowskian expansion (power series in Newton's constant $G$). This type of approach, in which the multipole moments can be viewed as the source moments (distinct from the radiative moments), was later investigated by Thorne \cite{Th80} and Blanchet and
Damour \cite{BD86} (we shall refer to \cite{BD86} as paper I).

The main result of paper I was to show that one can define starting from
the Thorne ``canonical'' linearized metric an explicit {\it algorithm}
for the computation of the exterior field up to any order in the
post-Minkowskian expansion.  The resulting metric represents the most
general post-Minkowskian solution of the vacuum equations outside the
source (up to a coordinate transformation).  It is parametrized by two
and only two sets of STF (source) multipole moments.  Following paper I we
shall call MPM (Multipolar-post-Minkowskian) this non-linear metric.  It
was also proved \cite{B87} that the re-expansion of the MPM metric at
future null infinity is consistent with the corresponding expansion
constructed by Bondi {\it et al} \cite{BBM62}.  This permits defining the radiative multipole moments within the framework of MPM metrics.

After dealing with the structure of the multipole expansion
\cite{Bo59,Th80,BD86,B87}, the next step is like in the linearized
theory to express the multipole moments as explicit integrals extending
over the source. At the leading order in a slow-motion expansion (speed of light $c\to\infty$), the multipole moments are given by the standard moments of the Newtonian mass and current densities in the source (see e.g.  \cite{Th80}).  At the first post-Newtonian (1PN) order, the source moments depend on the total stress-energy distribution of the matter fields {\it and} the gravitational field.  Because of the contribution of the gravitational field, the support of the total stress-energy distribution is
not spatially compact, and therefore the standard formulas used in the
linearized theory \cite{M62,CM71,CMM77,DI91b} do not apply.  If
nevertheless one tries to define the source multipole moments using these
formulas one obtains some expressions which are (typically) divergent
due to the behaviour of the integrals at spatial infinity.  This was the
approach followed initially by Epstein and Wagoner \cite{EW75} (see also
\cite{WagW76}) and generalized formally by Thorne \cite{Th80} to all
post-Newtonian orders.  The problem of divergencies becomes very
important starting at the 1.5PN order because of the appearance of
non-linear tails in the radiation field.  Indeed one can show \cite{B95}
that ignoring the divergent terms in the multipole moments as defined
previously is equivalent to neglecting the latter tail effects.
Recently this problem has been solved by Will and Wiseman \cite{WWi96}
who showed how to compute the tails within the Epstein-Wagoner-Thorne
approach by means of finite expressions for the multipole moments.

A satisfying derivation of the mass-type multipole moments of the source
at 1PN order was done in \cite{BD89} using a method of asymptotic
matching to the interior field of the source.  The moments were shown to
be actually compact-supported, and to agree with the moments of Epstein
and Wagoner modulo the formal discarding of infinite surface terms. The
current-type multipole moments at 1PN order were obtained in \cite{DI91a}
using a similar method of matching. At the 1PN order there is agreement
between the radiative moments (at infinity) and the source moments.
Only at the 1.5PN order do they start differing because of the
contribution of tails in the wave zone \cite{BD92}.

To 2PN order the multipole moments were obtained using a matching by
Blanchet \cite{B95} (we shall refer to \cite{B95} as paper II), with a
rather transparent result which seemed to be amenable to generalization.  Namely, it was found that the expressions of the multipole moments depend on the stress-energy (pseudo-)tensor of the matter and gravitational fields in the same way as would be obtained by using incorrectly the formulas valid for compact-supported sources (i.e. \`a la Epstein-Wagoner-Thorne), {\it but} that the multipole moments are endowed with a finite part operation based on analytic continuation which makes them to be perfectly well-defined mathematically. The latter finite part operation was found to be the same as used in the construction of the exterior field in paper I; it was carried in paper II all the way from the definition of the exterior field to the final expression of the moments.

In this paper we shall basically generalize the result of paper II.
We shall not perform a matching ``order
by order'' as was done in paper II, but rather construct directly the
multipole expansion generated by the total stress-energy
pseudo-tensor of the source.  This entails finding a formula for the
multipole expansion generated by a non-compact-supported source.
The multipole moments we shall obtain are valid for
slowly-moving sources but to all orders in the slow-motion (post-Newtonian)
parameter. Of course, the lowest-order post-Newtonian terms in these
general expressions agree with the previous results
obtained in \cite{BD89,DI91a,B95}.  On the other hand, we have agreement
in the limit of linearized gravity with the result of \cite{DI91b}.

The multipole moments we obtain are fully consistent with the construction of MPM metrics in paper I as they parametrize the so-called ``particular''
linearized metric of paper I (which differs from the ``canonical''
metric by a gauge transformation). In a sense the present paper realizes
a complete matching of the external field of paper I to a slowly-moving
source.  Note that the multipole moments, though allowing correctly for
all the nonlinearities in the near-zone field, parametrize the {\it
linearized} (exterior) metric which needs to be iterated in order to
obtain the radiative moments at infinity (see paper I and \cite{B87}).
Thus the present results have to be combined with what we know on the
relation between the source multipole moments and the radiative ones
(see e.g.  \cite{BD92,B97quad} and Section VI).

Computing the moments to very high post-Newtonian order is part of a
research program accompanying the development of the gravitational-wave
detectors LIGO and VIRGO. Its motivation comes from the fact that in
the case of the radiation emitted by compact binary systems an extremely
precise prediction from general relativity will be necessary in order to
extract the full potential information contained in the signal.  By
application of paper II the theoretical prediction for binary systems
was obtained at the 2PN order \cite{BDI95} (and subsequently at 2.5PN
order \cite{B96}). Will and Wiseman \cite{WWi96} obtained independently
the result at 2PN order by application of their improved Epstein-Wagoner
formalism.  The work at the 3PN order \cite{BIJ98} will rely in large
part on the general expressions of the multipole moments as derived in
the present paper.

The plan of the paper (besides this introduction and two appendices) is as
follows.  In Section II we recall some material from paper I and state
our basic assumptions.  In Section III we derive the expression of the
multipole expansion of the gravitational field valid outside a general
(slowly-moving) source. Re-writing the multipole expansion in a
different form we obtain in Section IV the ``linearized'' metric (i.e.
a solution of the linearized field equations) which is at the basis of
the post-Minkowskian iteration of paper I.  Decomposing the linearized
metric into irreducible STF tensors yields in Section V the general
expressions of the moments.  The paper ends with a discussion on
the link between the source moments and the radiative moments (Section
VI).

\section{Recalls and basic assumptions} \label{sec:2}

The assumptions on which the present investigation is based are twofold.
First we have the assumptions sustaining the construction of {\it
vacuum} metrics by means of the Multipolar-Post-Minkowskian (MPM) method
of paper I \cite{BD86}.  The so-called MPM metrics aim at describing the
gravitational field in the region {\it exterior} to a general isolated
system. Second we supplement the MPM method by other assumptions concerning the metric inside the isolated system. Essentially we assume that the metric is everywhere regular (smooth), and admits inside the system a post-Newtonian expansion which matches in the exterior to the vacuum MPM metric.  The matching is understood in the usual sense of matching of asymptotic expansions. Physically the formalism is restricted to slowly-moving
sources, whose typical internal velocities define a small post-Newtonian parameter $\sim v/c$.

\subsection{Recalls concerning the exterior field} \label{sec:2.A}

The Multipolar-Post-Minkowskian ``exterior'' metric of paper I is
defined in the open domain $I\!\!R^3_* \times I\!\!R$ (where
$I\!\!R^3_*=I\!\!R^3-\{ {\bf 0}\}$), i.e.  in $I\!\!R^4$ deprived from
the spatial origin $r \equiv |{\bf x}| = 0$.  We denote it by $h_{\rm
ext}^{\mu\nu} \equiv \sqrt{-g_{\rm ext}}\, g_{\rm ext}^{\mu\nu} -
\eta^{\mu\nu}$, where $g_{\rm ext}^{\mu\nu}$ is the inverse and $g_{\rm
ext}$ the determinant of the usual covariant metric, and where
$\eta^{\mu\nu} =\eta_{\mu\nu} = {\rm diag} (-1,1,1,1)$ is the Minkowski
metric.  The MPM metric $h_{\rm ext}^{\mu\nu}$ is in the form of a
formal (infinite) post-Minkowskian expansion,

\begin{equation}
h_{\rm ext}^{\mu\nu}=Gh_1^{\mu\nu}+G^2h_2^{\mu\nu}+...+G^nh_n^{\mu\nu}
+...   \label{eq:2.1}
\end{equation}
($G$ is Newton's constant), which is such that all the coefficients of
the $G^n$'s admit a {\it finite} multipolar expansion in symmetric and
trace-free (STF) products of unit vectors ${\bf n}={\bf x}/r$, i.e.
${\hat n}_L \equiv$ STF$\{ n_L\}$ where $n_L=n_{i_1}n_{i_2}...n_{i_l}$
(with $L=i_1i_2...i_l$ a multi-index with $l$ indices;  see \cite{N} for
our notation).  The decomposition on the tensors ${\hat n}_L$ is
equivalent to the usual decomposition on the basis of spherical
harmonics.  Thus, $\forall n \in I\!\!N$,

\begin{equation}
h_n^{\mu\nu}({\bf x},t) = \sum_{l=0}^{l_{\rm max}} {\hat n}_L(\theta ,\phi)
 ~{}_L h_n^{\mu\nu}(r,t) \ .\label{eq:2.2}
\end{equation}
The expansion is considered to be finite at any given order $n$ because
in practical computations we are doing the construction of the metric
for each multipolar pieces separately.  Note, though, that $l_{\rm max}$
tends to infinity as $n \to \infty$.  It is hoped that at the end of the
construction it is possible to take the limit of an infinite number of
multipole contributions.

The MPM metric satisfies the Einstein vacuum equations everywhere
except at the origin $r=0$ [i.e. $\forall ({\bf x},t) \in I\!\!R^3_* \times
I\!\!R$].  Using harmonic coordinates, this means

\begin{eqnarray}
\partial_\nu h_{\rm ext}^{\mu\nu} &=& 0 \ , \label{eq:2.3} \\
\Box h_{\rm ext}^{\mu\nu} &=& \Lambda_{\rm ext}^{\mu\nu} \ ,\label{eq:2.4}
\end{eqnarray}
where the box symbol denotes the flat d'Alembertian operator
$\Box = \eta^{\rho\sigma} \partial_\rho \partial_\sigma$ (with
$\partial_\rho = \partial /\partial x^\rho$), 
and where $\Lambda_{\rm ext}^{\mu\nu}\equiv \Lambda^{\mu\nu}(h_{\rm ext})$  
is a gravitational source term, whose support extends out to infinity,
and which encompasses all the nonlinearities, quadratic at least, of the
field equations [$\Lambda^{\mu\nu}(h)$ depends on $h$ and its space-time
derivatives $\partial h$ and $\partial^2 h$].  The relation between
$\Lambda^{\mu\nu}(h)$ and the Landau-Lifshitz \cite{LL} pseudo-tensor is
$\Lambda^{\mu\nu} = {16\pi G} |g| t^{\mu\nu}_{\rm LL}/c^4 +
\partial_\rho h^{\mu\sigma} \partial_\sigma h^{\nu\rho} - h^{\rho\sigma}
\partial^2_{\rho\sigma} h^{\mu\nu}$.  By the harmonic--coordinate
condition (\ref{eq:2.3}) we have, $\forall ({\bf x},t) \in I\!\!R^3_*
\times I\!\!R$,

\begin{equation}
 \partial_\nu \Lambda_{\rm ext}^{\mu\nu} = 0 \ . \label{eq:2.5}
\end{equation} 

Following paper I we further assume that the metric is stationary in a
neighbourhood of past time-like infinity, i.e.  that there exists a
finite instant $-{\cal T}$ in the past such that

\begin{equation}
t \leq -{\cal T} \quad \Rightarrow \quad \partial /\partial t ~h_{\rm
ext}^{\mu\nu} ({\bf x},t) = 0 \ .  \label{eq:2.6}
\end{equation} 
This assumption permits to implement in a simple way the condition
of no-incoming radiation, and, more technically, to avoid any problem of
divergence at infinity of retarded integrals of non-linear
source terms (with non-compact support).  The
assumption (2.6) could presumably be weakened in order to allow for some
always radiating matter systems.  We assume also that the metric is
asymptotically Minkowskian in the past, in the sense that

\begin{equation}
t \leq -{\cal T} \quad \Rightarrow \quad {\rm lim}_{~|{\bf x}|\to \infty}
~h_{\rm ext}^{\mu\nu}({\bf x})= 0 \ . \label{eq:2.7}
\end{equation}

By replacing the MPM metric (\ref{eq:2.1})-(\ref{eq:2.2}) into the field
equations (\ref{eq:2.3})-(\ref{eq:2.4}) and equating the coefficients of
the $G^n$'s on both sides we obtain an infinite set of perturbation
equations to be satisfied by all the $h_ n^{\mu\nu}$'s:  $\forall n \geq
1$,

\begin{eqnarray}
 \partial_\nu h_n^{\mu\nu} &=& 0 \ , \label{eq:2.8} \\ 
\Box h_n^{\mu\nu}
 &=& N_n^{\mu\nu} \ .\label{eq:2.9}
\end{eqnarray}
For $n=1$ we have $N_1\equiv 0$.  To any order $n \geq 2$ the source
$N_n$ is known from the previous iterations.  For instance, if
$\Lambda(h)=N_2(h)+O(h^3)$ then $N_2\equiv N_2(h_1)$.  The solution
$h_{\rm ext}^{\mu\nu}$ represents the infinite sum of the solutions of
the perturbation equations (\ref{eq:2.8})-(\ref{eq:2.9}) to every
post-Minkowskian orders.  Damour and Schmidt \cite{DS90} have proved
that the MPM expansion is ``reliable'', in the sense that it is possible
to construct smooth one-parameter families of solutions of the vacuum
equations whose Taylor expansion when $G \to 0$ belongs to the
class of MPM metrics.

The construction of the MPM metric proceeds iteratively
starting from any linearized metric $h_1$, solution in $I\!\!R^3_*
\times I\!\!R$ of the ``linearized'' vacuum equations [i.e.
(\ref{eq:2.8})-(\ref{eq:2.9}) where $n=1$ and $N_1\equiv 0$] and the
condition of retarded potentials.  The most general linearized metric can be written as an explicit multipolar expansion parametrized by a set of STF time-varying multipole moments.  This is
the so-called ``particular'' metric of paper I we recall in
(\ref{eq:4.9})-(\ref{eq:4.13}) below.  The ``particular'' metric
differs from the ``canonical'' metric of Thorne \cite{Th80} by a gauge
transformation.

As $h_1$ is in the form of a multipole expansion it is singular
at $r=0$, and so is $N_2$, and then successively $h_2$, ..., $N_n$.
To deal with this problem we apply when solving (2.9) the standard
retarded integral $\Box^{-1}_R$ on the product of $N_n$ and a factor
$(r/r_0)^B$ where $B$ is a complex number and $r_0$ a constant with the dimension of a length. The $B$-dependent retarded
integral $\Box^{-1}_R [(r/r_0)^B N_n]$ then defines a function of $B$
which is valid (by analytic continuation) for all values of $B$ in
$I\!\!\!C-Z\!\!\!Z$.  Near the value $B=0$ this function
admits a Laurent expansion, and it was shown in paper I that the {\it
finite part} at $B=0$ (denoted by ${\rm FP}_{B=0}$) of the Laurent
expansion when $B \to 0$, i.e.  the coefficient of the zeroth power of
$B$, is a solution in $I\!\!R^3_* \times I\!\!R$ of the d'Alembertian
equation with source $N_n$.  That is, we pose

\begin{equation}
u_n^{\mu\nu} = {\rm FP}_{B=0}\,\Box^{-1}_R [(r/r_0)^B N_n^{\mu\nu}]
     \ ,\label{eq:2.10}
\end{equation}
where $\Box^{-1}_R$ denotes the retarded integral

\begin{equation}
(\Box^{-1}_R N)({\bf x},t) \equiv - {1\over 4\pi}
\int {d^3{\bf y}\over |{\bf x}-{\bf y}|} 
N( {\bf y}, t - |{\bf x} -{\bf y}|/c )\ . \label{eq:2.11} 
\end{equation}
Then we have, $\forall ({\bf x},t) \in I\!\!R^3_* \times I\!\!R$,
$\Box u_n^{\mu\nu} = N_n^{\mu\nu}$. 

In order to satisfy the gauge condition (\ref{eq:2.8}) it is necessary
to add to the particular solution $u_n$ a certain homogeneous solution
$v_n$ whose divergence cancels out the divergence of $u_n$:
$\partial_\nu v_n^{\mu\nu}=-\partial_\nu u_n^{\mu\nu}$.  Following the
algorithm proposed in paper I we must decompose $\partial_\nu
u_n^{\mu\nu}$ into four STF tensors $A_L$, $B_L$, $C_L$ and $D_L$ and
apply the equations (4.13) in paper I.  A slightly modified algorithm,
which is more convenient for our purpose, has been
defined in the equations (2.12) of \cite{B97quad}.  In Appendix B we
recall the definition of the modified algorithm for $v_n$ in terms of
the tensors $A_L$, $B_L$, $C_L$ and $D_L$.  Thus the solution of
(\ref{eq:2.8})-(\ref{eq:2.9}) reads ($\forall n \geq 2$)

\begin{equation}
h_n^{\mu\nu} = u_n^{\mu\nu} + v_n^{\mu\nu} \ . \label{eq:2.12}
\end{equation}

When starting from the ``particular'' linearized metric $h_{\rm part.1}$ [see (\ref{eq:4.11})-(\ref{eq:4.13})], the previous algorithm for the
construction of the MPM metric generates in fact the {\it most general} solution of the vacuum equations under the MPM assumptions. When starting from the ``canonical'' metric $h_{\rm can.1}$ the MPM algorithm still generates the most general solution but modulo a coordinate transformation. See Theorems 4.2 and 4.5 in paper I. Crucial to the construction of the MPM metric is the knowledge of the general structure of the singularity when $r \to 0$ of each of the post-Minkowskian coefficients $h_n$, which permits to
legitimate the application of the operator FP$\Box^{-1}_R$ at each post-Minkowskian order. The following result was proved in paper I.

\medskip\noindent
{\it Theorem}. The
expansion when $r \to 0$ of any post-Minkowskian coefficient
$h_n$ reads, $\forall ({\bf x},t) \in
I\!\!R^3_* \times I\!\!R$ and $\forall N \in I\!\!N$,

\begin{equation}
 h_n({\bf x},t) = \sum_{a\leq N} {\hat n}_L r^a (\ln r)^p
{}_LF_{n,a,p}(t)+R_{n,N}({\bf x},t)  \ , \label{eq:2.13}
\end{equation} 
where $a\in Z\!\!\!Z$ with $-a_0 \leq a \leq N$ (and $a_0\in I\!\!N$),
$p \in I\!\!N$ with $p \leq n-1$, and still $l \leq l_{\rm max}$ [see (5.4) in paper I]. Both $a_0$ and $l_{\rm max}$ depend on $n$, and tend to infinity when$n \to \infty$, as does the maximal power of the logarithms, $p_{\rm max}=n-1$. [The logarithms in (2.13) should really be $\ln (r/r_0)$ but we include the constants $\ln r_0$ into the definition of ${}_LF_{n,a,p}$.] The functions ${}_LF_{n,a,p}(t)$ are smooth ($C^\infty$)
functions of time [starting with $C^\infty$ multipole moments at the
linearized level], and constant in the remote past: ${}_LF_{n,a,p}(t) = {\rm const}$ when $t \leq -{\cal T}$. They are given by some complicated non-linear functionals of the (source) multipole moments parametrizing the linear metric. The remainder $R_{n,N}({\bf x},t)$ belongs to the so-called class of functions $O^N(r^N)$ introduced in paper I, i.e.  basically all its time derivatives are zero in the past ($t \leq -{\cal T}$), are of class
$C^N(I\!\!R^4)$, and are of order $O(r^N)$ when $r \to 0$ (we refer to paper I for full mathematical details).

To prove this theorem, one assumes as an induction hypothesis that $N_n$
admits the same type of structure as (\ref{eq:2.13}), with the only
exception that $p_{\rm max}=n-2$.  Then one shows that applying
FP$\Box_R^{-1}$ on each of the terms composing (\ref{eq:2.13}) makes
sense (by analytic continuation), and that this is a stable operation in
the sense that the ``solution'' $u_n$, and then $h_n=u_n+v_n$, admits the same structure as the ``source'' $N_n$, with merely an increase by one unit
of the maximal power of the logarithms.  Since (\ref{eq:2.13}) is
manifestly correct for $h_1$ if $p_{\max}=0$, one concludes that (2.13)
is correct $(\forall n$) with the result that $p_{\rm max}=n-2$ for
$N_n$ and $p_{\rm max}=n-1$ for $h_n$.
 
\subsection{Assumptions concerning the inner field} \label{sec:2.B}

We now address the problem of the gravitational field inside an actual
isolated system, described by a stress-energy tensor $T^{\mu\nu} ({\bf
x},t)$ in some coordinate system $({\bf x},t)$, with spatial origin
$r\equiv |{\bf x}| =0$ located within the system. We suppose that $T^{\mu\nu}$ is stationary in the past, and that its 
support is spatially compact, $T^{\mu\nu} ({\bf x},t) =
0$ when $r >d$. The metric $h^{\mu\nu} \equiv \sqrt{-g}\, g^{\mu\nu}
- \eta^{\mu\nu}$ satisfies the Einstein field equations throughout
$I\!\!R^4$, thus [using harmonic coordinates like in
(\ref{eq:2.3})]

\begin{eqnarray}
 \partial_\nu h^{\mu\nu} &=& 0 \ , \label{eq:2.14} \\ \Box h^{\mu\nu}
 &=& {16\pi G\over c^4} \tau^{\mu\nu} \ ,\label{eq:2.15}
\end{eqnarray}
where we have defined the effective stress-energy (pseudo-)tensor

\begin{equation}
   \tau^{\mu\nu} \equiv |g| T^{\mu\nu} + {c^4 \over 16\pi G}
   \Lambda^{\mu\nu}\ . \label{eq:2.16}
\end{equation} 
The first term in (\ref{eq:2.16}) is the matter source term, which is of
compact support.  The second term is the non-compact-supported
gravitational source term, which is the same as in
(\ref{eq:2.4}) but is computed with $h$ instead of $h_{\rm ext}$,
i.e.  $\Lambda^{\mu\nu}=\Lambda^{\mu\nu}(h)$.  From
(\ref{eq:2.14}) the divergence of the (relaxed) Einstein equation
(\ref{eq:2.15}) yields the equation of motion of the matter fields,

\begin{equation}
 \partial_\nu \tau^{\mu\nu} = 0 \ , \label{eq:2.17}
\end{equation} 
which is equivalent to the equation of conservation of $T^{\mu\nu}$ in
the covariant sense $(\nabla_\nu T^{\mu\nu} =0)$.  The retarded solution
of the system of equations (\ref{eq:2.14})-(\ref{eq:2.16}) is obtained
by simple inversion of the flat d'Alembertian operator,

\begin{equation}
 h^{\mu\nu} = {16\pi G\over c^4} \Box ^{-1}_{R} \tau^{\mu\nu}\ ,
 \label{eq:2.18}
\end{equation} 
where $\Box^{-1}_R$ is given by (\ref{eq:2.11}).
Since the pseudo-tensor $\tau^{\mu\nu}$ depends on $h$ itself and its
derivatives, the equation (\ref{eq:2.18}) [with the constraint equation
(\ref{eq:2.17})] should be viewed in fact as an integro-differential
equation equivalent to (\ref{eq:2.14})-(\ref{eq:2.16}) and the condition
of retarded potentials.

We now assume that $h^{\mu\nu}$, as given by (\ref{eq:2.18}) with
(\ref{eq:2.17}), safisfies certain mathematical properties complementing the
properties of the exterior MPM metrics.  Here we state our assumptions and comment on their adequacy in the case of the gravitational field.  Based on these assumptions we shall construct in Section III the multipole expansion of $h^{\mu\nu}$.

\bigskip \noindent
{\it Assumption} (i) (smoothness).  The field $h$ (skipping the
space-time indices) is a smooth function of the harmonic coordinates
(${\bf x},t$), namely 

\begin{equation}
h({\bf x},t)~\in~C^\infty(I\!\!R^4) \ . \label{eq:2.19}
\end{equation} 
In particular we assume no singularities
in the distribution of matter fields (no point particles or black
holes).  The formalism will be {\it a priori} valid only for continuous
matter distributions.  However, there are indications that the
formalism applies also to point particles modelling compact objects like
black holes (see \cite{BD92,BDI95,BIJ98}).  Simply one needs to
represent the matter fields by delta-functions and to use a
regularization in order to deal with the field near the point particles.

\bigskip \noindent
{\it Assumption} (ii) (multipole expansion).  $h$ admits a multipolar
expansion in the open domain exterior to the compact support of the
matter source $T^{\mu\nu}$, in the sense that it agrees there
with the MPM metric $h_{\rm ext}$ constructed in paper I.  Throughout
this paper, we denote the multipole expansion using the script letter
${\cal M}$. Thus, by definition,

\begin{equation}
{\cal M}(h) \equiv h_{\rm ext} \ . \label{eq:2.20}
\end{equation}
Let the exterior domain be $r > {\cal R}$, where the constant ${\cal R}$ is such that ${\cal R}> d$ (with $d$ the maximal radius of the compact
support of $T^{\mu\nu}$). Our assumption reads

\begin{equation}
r > {\cal R} \quad\Rightarrow\quad {\cal M}(h) = h \ . \label{eq:2.21}
\end{equation} 
Note that (\ref{eq:2.20}) simply means that we have given to $h_{\rm
ext}$ the {\it name} ${\cal M}(h)$ (multipole expansion of $h$);
thus ${\cal M}(h)$ is a solution of the vacuum equations valid
everywhere except at the spatial origin, i.e.  $\forall ({\bf x},t) \in
I\!\!R^3_* \times I\!\!R$.  By contrast (\ref{eq:2.21}) states that at
any field point located in the exterior of our physical system ($r >
{\cal R} > d$), ${\cal M}(h)$ agrees {\it numerically} with $h$.  Of
course, $h$ satisfies inside the system the non-vacuum field equations
and therefore differs from the vacuum solution ${\cal M}(h)$.  For
instance $h$ is smooth all-over the system [assumption (i)] while ${\cal
M}(h)$ is singular at $r=0$.

Summing up all the post-Minkowskian contributions given by
(\ref{eq:2.13}) let us write the theorem giving the structure of the
expansion when $r \to 0$ of ${\cal M}(h)$:  $\forall ({\bf x},t) \in
I\!\!R^3_* \times I\!\!R$, $\forall N \in I\!\!N$,

\begin{equation}
 {\cal M}(h)({\bf x},t) = \sum_{a\leq N} {\hat n}_L r^a (\ln r)^p
{}_LF_{a,p}(t)+R_N({\bf x},t)  \ , \label{eq:2.22}
\end{equation} 
where the functions ${}_LF_{a,p}$ and $R_N$ are of the type $\sum_n G^n
{}_LF_{n,a,p}$ and $\sum_n G^n R_{n,N}$.  Note that, in contrast with
(\ref{eq:2.13}), the summation integers $a$, $p$ and $l$ in
(\ref{eq:2.22}) take an infinite number of values:  $ a\in Z\!\!\!Z$ is
not bounded from below (i.e.  $-\infty \leq a \leq N$), and $p \in
I\!\!N$, $l \in I\!\!N$ are not bounded from above ($0 \leq p, l \leq
+\infty$).  We refer to the ``complete'' expansion including all values of
$a\in Z\!\!\!Z$ in (\ref{eq:2.22}) as the ``near-zone'' expansion ($r \to 0$) of the multipole expansion, and denote it with an overbar,

\begin{equation}
 \overline {{\cal M}(h)}({\bf x},t) = \sum {\hat n}_L
r^a (\ln r)^p {}_LF_{a,p}(t)\ . \label{eq:2.23}
\end{equation} 
This expansion should be considered in the sense of formal series, 
i.e. as an infinite set of coefficients of $r^a(\ln r)^p$.

\bigskip \noindent {\it Assumption} (iii) (post-Newtonian expansion).
At any fixed space-time position (${\bf x},t$), $h$ admits an asymptotic
expansion when the speed of light $c \to +\infty$ along the basis of
functions $c^{-i}(\ln c)^q$ where $i,q \in I\!\!N$ ($h$ depends on $c$ as a solution of the field equations).
Thus, $\forall N \in I\!\!N$,

\begin{equation}
 h({\bf x},t,c) = \sum_{i\leq N} c^{-i} (\ln c)^q \sigma_{i,q} ({\bf
 x},t)+T_N({\bf x},t,c) \ , \label{eq:2.24}
\end{equation} 
where the coefficients $\sigma_{i,q}$ are smooth functions of (${\bf
x},t$), and where the remainder $T_N$ is of order $O(1/c^N)$ when $c \to
\infty$ (non-uniformly in $|{\bf x}|$).
The corresponding expansion up to any order $i \in I\!\!N$
is
referred to as the post-Newtonian expansion of $h$ and denoted also with
an overbar (this abuse of notation being justified by the matching
below),

\begin{equation} 
\overline h ({\bf x},t,c) = \sum c^{-i} (\ln c)^q
\sigma_{i,q} ({\bf x},t) \ .  \label{eq:2.25} 
\end{equation} 
Like
(\ref{eq:2.23}) this expansion is to be viewed in the sense of formal
series.  Clearly the assumed structure (\ref{eq:2.24})-(\ref{eq:2.25})
of the post-Newtonian expansion is consistent with
(\ref{eq:2.22})-(\ref{eq:2.23}). Indeed it is known (paper I) that
the MPM exterior field, namely the multipole expansion ${\cal M}(h)$,
depends on the radial coordinate $r$ only through the ratio $r/c$ (when
the multipole moments are considered to be independent functions of
time).  Thus the near-zone re-expansion of the multipole expansion,
$\overline{{\cal M}(h)}$, can be viewed equivalently as an expansion
when $c \to \infty$, which should be necessarily of the type $\sum
c^{-i} (\ln c)^q$ [replacing $r$ by $r/c$ in (2.23)].  Our assumption
(\ref{eq:2.25}) means in fact that the multipole moments parametrizing
the linearized metric admit themselves, when expressed in terms of the
source parameters, the same type of expansion $\sum c^{-j} (\ln c)^s$. There
are many indications that the post-Newtonian expansion involves besides
the usual powers of $1/c$ some (powers of) logarithms of $c$ (see paper
I and references therein). As usual (because all retardations
$r/c$ associated with the propagation of gravity at the speed of light
tend to zero), the post-Newtonian expansion is valid only in a region
surrounding the source which is of small extent as compared to one
characteristic wavelength of the emitted radiation, i.e.  $\overline h$
constitutes a good approximation to $h$ only in the region $r \ll
\lambda$.

\bigskip \noindent
{\it Assumption} (iv) (matching).  A ``matching'' region around the
source exists, where the multipole expansion ${\cal M}(h)$ and the
post-Newtonian expansion $\overline h$ are simultaneously valid.  In
this region one expects $r > {\cal R}$ {\it and} $r \ll \lambda$ so that
${\cal R} \ll \lambda$, which implies since ${\cal R} > d$ that the
size of the source is $d \ll \lambda$ or equivalently that the typical
internal velocities within the source are $v \approx d c/\lambda \ll
c$.  Existence of the matching region means therefore {\it
slowly-moving} source.  In this region we have the numerical
equality [from (\ref{eq:2.21})]

\begin{equation}
{\cal R} < r \ll \lambda \quad\Rightarrow\quad {\cal M}(h) = \overline h
 \ . \label{eq:2.26}
\end{equation} 
We now transform (2.26) into a matching equation, i.e.  an equation
relating two mathematical expressions of the same nature, by replacing in
the left side ${\cal M}(h)$ by its near-zone expansion $\overline{{\cal
M}(h)}$ as given by (\ref{eq:2.23}), and in the right side $\overline h$
by its multipole expansion ${\cal M}(\overline h)$ obtained from
(\ref{eq:2.25}) by substituting each of the coefficients $\sigma_{i,q}$
by their multipole expansion ${\cal M}(\sigma_{i,q})$.
Actually we have not defined what we mean by ${\cal M}(\sigma_{i,q})$, as
this would necessitate to perform a ``multipolar-post-Newtonian'' iteration
of the vacuum equations, analogous to the MPM iteration of paper I.
Simply we assume the existence of each ${\cal M}(\sigma_{i,q})$,
and we shall determine its structure (2.30) as a consequence of the matching. Similarly to (\ref{eq:2.21}) the multipole expansion of $\overline h$ satisfies ({\it term by term} in the post-Newtonian expansion)

\begin{equation}
r > {\cal R} \quad\Rightarrow\quad {\overline h} = {\cal M}({\overline h})
\ . \label{eq:2.27}
\end{equation}
Now the matching equation associated with (\ref{eq:2.26}) reads

\begin{equation}
\overline {{\cal M}(h)}  = {\cal M}(\overline h)  \ . \label{eq:2.28}
\end{equation} 
Formally this equation is true  ``everywhere'', in the sense that it represents an infinite set of {\it functional} equalities [valid $\forall ({\bf x},t) \in I\!\!R^3_* \times I\!\!R$] between each of the coefficients of the series in both sides.  Of course, the two series should be re-arranged as expansions along the same basis of functions, i.e. either $r^a (\ln r)^p$ or $c^{-i} (\ln c)^q$.  Note that the matching equation shows that the right side of (2.23) represents equivalently the ``near-zone'' expansion ($r\to 0$) of ${\cal M}(h)$ and the ``far-zone'' expansion $(r\to +\infty)$ of $\overline h$. From (\ref{eq:2.28}) one deduces that the
structure of the functions ${}_LF_{a,p}$ in (\ref{eq:2.23}) in terms of the independent variable $c$ is

\begin{equation} {}_LF_{a,p} (t,c)=\sum c^{-i}
(\ln c)^q {}_LG_{a,p,i,q} (t) \ , \label{eq:2.29}
\end{equation}
{\it and} that the structure of each multipole expansion ${\cal
M}(\sigma_{i,q})$ reads

\begin{equation}
 {\cal M}(\sigma_{i,q})({\bf x},t) = \sum {\hat n}_L r^a (\ln r)^p
{}_LG_{a,p,i,q}(t) \ . \label{eq:2.30}
\end{equation} 
It is important to realize that although the matching is performed in
the (exterior) near-zone where the post-Newtonian
expansion is valid (because of the small retardation $r/c$), the
multipole expansion (\ref{eq:2.30}) of the post-Newtonian coefficients
is valid wherever $r>{\cal R}$, and {\it in particular} when $r \to
\infty$.  Actually (2.30) represents the far-zone expansion of
$\sigma_{i,q}$ (at spatial infinity).  Typically ${\cal M}(\sigma_{i,q})$
blows up in the far-zone, as is clear from the positive powers of $r$ in
(\ref{eq:2.30}).  But this is not a problem, as this simply reflects the
fact that the post-Newtonian expansion is not valid in a neighbourhood
of infinity (where it would constitute a very poor approximation of
$h$).

The assumption (iv) is an application of the method of matched
asymptotic expansions. In the present context it permits to ``anchor'' the
multipole expansion to the field inside the actual source.  Matched
asymptotic expansions were used in general relativity originally for
dealing with radiation reaction problems \cite{Bu71,BuTh70}, and, within
MPM expansions, in order to find the expression of the multipole moments
with increasing post-Newtonian precision \cite{BD89,DI91a,B95,B96}.
Each time, the matching was found consistent in the sense that the
satisfaction of the equation (2.28) determined all the desired
information at the required order.

The above assumptions (i-iv) are natural complements of the MPM
framework; they should apply generically to the field generated by an
isolated system.  However let us stress that we have made three physical
restrictions on the system:  that it be stationary in the remote past
(before the instant $-{\cal T}$), slowly-moving (existence of a small
post-Newtonian parameter $\sim v/c$), and without singularities (no
point particles or black holes).

\section{Multipole expansion of the gravitational field} \label{sec:3}

Our assumptions being stated, we start our investigation. Notice that the assumptions (i-iv) have been written for the field $h$, but we can readily prove with the help of the field equation (\ref{eq:2.15}) that they apply as well to the stress-energy pseudotensor $\tau$.  In particular since ${\cal M}(\Lambda)=16\pi G/c^4 {\cal M}(\tau)$ [because the matter stress-energy tensor $T$ has a compact support] we see that ${\cal M}(\Lambda)$, which is nothing but the MPM source $\Lambda_{\rm ext}$ in (\ref{eq:2.4}), admits exactly the same structure (\ref{eq:2.22}) as ${\cal M}(h)$. Therefore we can apply on ${\cal M}(\Lambda)$ the finite part at $B=0$ of the operator $\Box_R^{-1}(r/r_0)^B$ whose definition was given in (\ref{eq:2.10})-(\ref{eq:2.11}). We have seen that since ${\cal M}(\Lambda)$ involves an infinite sum of post-Minkowskian contributions, the summation in (\ref{eq:2.22}) is infinite (notably $-\infty \leq a \leq N$).  Thus we mean by ${\rm FP} \Box_R^{-1}{\cal M}(\Lambda)$ the infinite sum of the ${\rm FP}
\Box_R^{-1}$'s acting on each separate terms composing ${\cal M}(\Lambda)$.  Recall that we are working in a context of approximate solutions, which are constructed only up to a given (though arbitrary) post-Minkowskian order, and thus involve only a finite number of separate contributions.
Here we assume that we can consider the {\it formal} series of
approximations, but we do not control the precise mathematical nature
of this series (see, however, \cite{DS90}).

We can perform our derivation restricting our
attention to the case where ${\cal M}(\Lambda)$ is not only constant in the
past but {\it zero} in the past ($t \leq -{\cal T}$).  Indeed we can check
{\it a posteriori} that the derivation can be redone straightforwardly
in the case of the constant terms in ${\cal M}(\Lambda)$ by simply using the
Poisson operator $\Delta^{-1}$ instead of the retarded integral
$\Box^{-1}_R$.  Thus the constant terms can be added to the result at
the end with no modification except that $\Box^{-1}_R$ becomes
$\Delta^{-1}$ when acting on such terms.

From paper I the $B$-dependent retarded integral $\Box_R^{-1}[(r/r_0)^B{\cal
M}(\Lambda)]$ admits a Laurent expansion when $B \to 0$ whose finite
part solves the d'Alembertian equation with source ${\cal M}(\Lambda)$:
$\forall ({\bf x},t) \in I\!\!R^3_* \times I\!\!R$,

\begin{equation}
\Box \left\{{\rm FP}_{B=0}\, \Box^{-1}_R [ \tilde r^B {\cal
M}(\Lambda)] \right\} = {\cal M}(\Lambda) \ . \label{eq:3.1}
\end{equation} 
From now on we pose $\tilde r =r/r_0$. Beyond the finite part all possible multiple poles $\sim B^{-i}$ exist {\it a priori} ($\forall i \in I\!\!N$;  indeed $i \leq i_{\rm max}$ to any post-Minkowskian order $n$ but $i_{\rm max} \to \infty$ when $n \to \infty$).  

We want to compute the multipole expansion of the field, namely ${\cal M}(h)\equiv 16\pi G/c^4 {\cal M}(\Box^{-1}_R \tau)$. To do this we remark that because the multipole expansion satisfies $\Box{\cal M}(h)={\cal
M}(\Lambda)$ (all-over $I\!\!R^3_* \times I\!\!R$), we have for any $B$
the relation $\Box [\tilde r^B {\cal M}(h)]=\tilde r^B \{ {\cal M} (\Lambda)+2Br^{-1} \partial_r {\cal M}(h)+B(B+1)r^{-2} {\cal M}(h) \}$, where $\partial_r \equiv n_i\partial_i$.  Applying on both sides of this relation the retarded integral $\Box^{-1}_R$ and considering the finite part at $B=0$ yields

\begin{eqnarray}
{\cal M}(h) &=& {\rm FP}_{B=0}\, \Box^{-1}_R [ \tilde r^B
{\cal M}(\Lambda)] \nonumber \\
&+& {\rm FP}_{B=0}\, \Box^{-1}_R [ \tilde r^B \{ 2Br^{-1} \partial_r
{\cal M}(h) +B(B+1)r^{-2} {\cal M}(h) \} ] \ .   \label{eq:3.2}
\end{eqnarray} 
The first term will ensure that ${\cal M}(h)$ is a particular solution
of the correct equation $\Box{\cal M}(h)={\cal M}(\Lambda)$ [see
(\ref{eq:3.1})].  This particular solution is already in the
appropriate form, thus we concentrate our attention on the second term,
which constitutes an homogeneous solution of the wave equation,
as will follow from the fact that this second term in (\ref{eq:3.9}) involves an explicit factor $B$ (so it appears like a residue rather than a finite part). Because of this factor $B$ we see that the contribution of
any term in (\ref{eq:2.22}) which is regular at $r=0$ will be exactly zero.
This is in particular the case of the remainder $R_N$ in (\ref{eq:2.22}) which will give zero, and since this is true $\forall N \in I\!\!N$ we conclude that ${\cal M}(\Lambda)$ can be replaced by the infinite expansion $\overline {{\cal M}(\Lambda)}$ given by (\ref{eq:2.23}).  Thus, writing the retarded integral in the right side of (\ref{eq:2.23}) in the form (\ref{eq:2.11}),

\begin{eqnarray}
{\cal M}(h) &=& {\rm FP}_{B=0}\, \Box^{-1}_R [ \tilde r^B
{\cal M}(\Lambda)] \nonumber \\
&-& {1\over 4\pi}
{\rm FP}_{B=0} \int {d^3{\bf y} |{\tilde {\bf y}}|^B \over |{\bf x}-{\bf y}|}
[ 2Br^{-1} \partial_r
\overline{{\cal M}(h)} +B(B+1)r^{-2} \overline{{\cal M}(h)}] 
( {\bf y}, t - |{\bf x} -{\bf y}|/c )\ , \label{eq:3.3}
\end{eqnarray}
where we denote $\tilde{\bf y}\equiv {\bf y}/r_0$. The factor $B$ shows that the only contribution to the triple
integral is due to a (simple) pole at $B=0$, which in turn comes only
from the integration on an infinitesimal neighbourhood of the spatial
origin, $|{\bf y}| < \varepsilon$ where $\varepsilon$ is an
arbitrary small number ($0 < \varepsilon \ll |{\bf x}|$ say). [As
we have assumed that the functions ${}_LF_{a,p}$ are zero in the past
(we add the constant parts at the end) there is no problem at the upper
bound $|{\bf y}| \to \infty$.] Then it is legitimate to replace the
integrand in (\ref{eq:3.3}) by its Taylor expansion when $|{\bf y}| \to
0$. The Taylor expansion of any $K(t-|{\bf x} -{\bf y}|/c)/|{\bf
x}-{\bf y}|$ is given by $\sum (-)^l y_L \partial_L\{K(t-r/c)/r\}/{l !}$
(see \cite{N} for the notation).  In this way we obtain

\begin{equation}
{\cal M}(h) = {\rm FP}_{B=0}\, \Box^{-1}_R [ \tilde r^B {\cal M}(\Lambda)] 
  - {4G \over c^4} \sum^{+\infty}_{l=0}
 {(-)^l\over l!} \partial_L \left\{ {1\over r} {\cal K}_L (t-r/c)
 \right\} \ , \label{eq:3.4}
\end{equation}
where the ``multipole
moments'' ${\cal K}_L(u)$ are given explicitely by ($u\equiv t-r/c$)

\begin{equation}
 {\cal K}_L (u) = {c^4\over 16\pi G}{\rm FP}_{B=0}  \int_{|{\bf y}|  < \varepsilon} d^3 {\bf y} |\tilde{\bf y}|^B y_L [ 2Br^{-1} \partial_r
 \overline{{\cal M}(h)} +B(B+1)r^{-2} \overline{{\cal M}(h)}]
({\bf y}, u)\ .
 \label{eq:3.5}
\end{equation}
 
Next we perform a sequence of transformations of ${\cal K}_L (u)$.
By (\ref{eq:2.23}), the structure of $\overline{{\cal M}(\Lambda)}({\bf
y}, u)$ is that of a series of terms of the type $\sum {\hat n}_L |{\bf
y}|^a (\ln |\tilde {\bf y}|)^p {}_LF_{a,p} (u)$.  This shows that after
integrating over the angles ${\cal K}_L(u)$ is made of a series of terms of the type ${\rm FP}_{B=0} B\int_0^\varepsilon d|{\bf y}| |\tilde{\bf y}|^B |{\bf y}|^{2+l+a}(\ln |\tilde {\bf y}|)^p$ times a function of $u$. The
latter radial integrals are defined by analytic continuation in $B$.
Now the point is that the {\it complete} radial integral extending from
$0$ to $+\infty$, i.e. $\int_0^{+\infty} d|{\bf y}| |\tilde{\bf y}|^B |{\bf
y}|^{2+l+a}(\ln |\tilde{\bf y}|)^p$, is rigorously {\it zero} by analytic
continuation, for all values of $B \in I\!\!\!C$. [We repeat the reasoning already made in paper II: the integral can be split into the sum of two integrals, namely $(d/dB)^p \int_0^{\cal A} d|{\bf y}||\tilde{\bf y}|^B  |{\bf y}|^{2+l+a}$ and $(d/dB)^p \int_{\cal A}^{+\infty} d|{\bf y}| |\tilde{\bf y}|^B |{\bf y}|^{2+l+a}$, where $\cal A$ is some positive constant. When the real part of $B$ is a large positive number, the first integral reads $(d/dB)^p\{\tilde {\cal A}^B {\cal A}^{3+l+a}/(B+3+l+a)\}$; and when the real part of $B$ is a large negative number, the second integral is equal to the opposite $-(d/dB)^p\{ \tilde{\cal A}^B {\cal A}^{3+l+a}/(B+3+l+a)\}$. These expressions represent the unique analytic continuations of the two integrals for all values of $B$ except $-(3+l+a)$. Now $\int_0^{+\infty} d|{\bf y}| |\tilde {\bf y}|^B |{\bf
y}|^{2+l+a}(\ln |\tilde{\bf y}|)^p$ is defined by analytic continuation to be the sum of the analytic continuations of the two separate integrals, and is therefore identically zero ($\forall B \in I\!\!\!C$).] Thus all the previous terms can be equivalently written as $-{\rm FP}_{B=0} B \int_\varepsilon^{+\infty} d|{\bf y}||\tilde {\bf y}|^B |{\bf y}|^{2+l+a} (\ln |\tilde{\bf y}|)^p$.  Furthermore, using now the presence of the explicit factor $B$, we see that the only contribution to the finite part at $B=0$ comes from an arbitrary neighbourhood of the upper bound $|{\bf y|}=+\infty$. In this paper it is sufficient to consider as neighbourhood of infinity the domain $|{\bf y}|>{\cal R}$, where ${\cal R}$ denotes the radius giving the limit of validity of multipole expansions.  Therefore ${\cal K}_L(u)$ is given as well by a series of terms of the type $-{\rm
FP}_{B=0} B\int_{\cal R}^{+\infty} d|{\bf y}| |\tilde {\bf y}|^B |{\bf y}|^{2+l+a}(\ln |\tilde{\bf y}|)^p$, which shows that (\ref{eq:3.5}) is in fact equivalent to

\begin{equation}
 {\cal K}_L (u) = - {c^4\over 16\pi G}{\rm FP}_{B=0}  \int_{|{\bf y}|>{\cal R}} d^3 {\bf y} |\tilde{\bf y}|^B  y_L [ 2Br^{-1} \partial_r
\overline{{\cal M}(h)} +B(B+1) r^{-2} \overline{{\cal M}(h)}]
({\bf y}, u)\ .
 \label{eq:3.6}
\end{equation}
Note the minus sign with respect to the previous expression
(\ref{eq:3.5}).  The next step is to employ our assumption (iv) of
consistent matching between the post-Newtonian and multipole expansions,
according to which one can commute the order of taking the
post-Newtonian and multipole expansions:  by (\ref{eq:2.28}) [or, rather,
$\Box$ of (2.28)], $\overline{{\cal M}(\Lambda)}$ and ${\cal
M}(\overline \Lambda)$ are functionally equal (term by term), so
(\ref{eq:3.6}) can be re-written as

\begin{equation}
 {\cal K}_L (u) = - {c^4\over 16\pi G}{\rm FP}_{B=0} \int_{|{\bf y}|>{\cal R}} d^3 {\bf y} |\tilde {\bf y}|^B y_L [ 2Br^{-1} \partial_r
{\cal M}(\overline{h}) +B(B+1)r^{-2} {\cal M}(\overline{h})]
({\bf y}, u)\ .
 \label{eq:3.7}
\end{equation}
But now in the region $|{\bf y}|>\cal R$ one can replace the multipole
expansion by the function itself [see (\ref{eq:2.27})].
Thus,

\begin{equation}
 {\cal K}_L (u) = - {c^4\over 16\pi G}{\rm FP}_{B=0}  \int_{|{\bf y}|>
{\cal R}} d^3 {\bf y} |\tilde{\bf y}|^B y_L [ 2Br^{-1} \partial_r
\overline{h} +B(B+1)r^{-2} \overline{h}] ({\bf y}, u)\ .
 \label{eq:3.8}
\end{equation}
Finally we recall that by our assumption (i) the field $h$ and hence
the function $\Lambda$ and all the $\sigma_{i,q}$'s in the post-Newtonian expansion ${\overline \Lambda}$ are regular ($C^\infty$) functions on $I\!\!R^4$, and in particular near the origin $r \to 0$. Therefore we see that the range of integration in (\ref{eq:3.8}) can be harmlessly extended to the whole three-dimensional space. Indeed because of the factor $B$ the
contribution to the integral due to any ball of finite radius, for
instance the ball ${|{\bf y}| \leq {\cal R}}$, will be zero after taking
the finite part at $B=0$.  Thus we have 

\begin{equation}
 {\cal K}_L (u) = - {c^4\over 16\pi G}{\rm FP}_{B=0}  \int 
d^3 {\bf y} |\tilde {\bf y}|^B  y_L [ 2Br^{-1} \partial_r
\overline{h} +B(B+1)r^{-2} \overline{h}]
({\bf y}, u)\ . \label{eq:3.9}
\end{equation}
Since $\Box\overline
h=16\pi G~{\overline \tau}/c^4$ we can re-combine the terms in the
integrand and find the equivalent expression

\begin{equation}
{\cal K}_L (u) = {\rm FP}_{B=0} \int d^3 {\bf y} y_L \, \left[ \tilde r^B
{\overline \tau} - {c^4\over 16\pi G} \Box(\tilde r^B {\overline h})\right] ({\bf y}, u)\ .
\label{eq:3.10}
\end{equation} 

The point about (\ref{eq:3.9}) or (\ref{eq:3.10}) is that because of the factor $B$ the numerical values of the
multipole moments depend on the post-Newtonian expansion of the source as
integrated formally in a neighbourhood of (spatial) infinity.  This is
despite the fact that the post-Newtonian expansion is expected {\it a
priori} to be valid only near the origin. As we have shown here, this
seemingly contradictory result is possible thanks to the properties of
analytic continuation, which permit to jump from a ``near-zone''
integration range in (\ref{eq:3.5}) to the ``far-zone'' in
(\ref{eq:3.6}). However we now remark that the contributions in the complete
multipole expansion (involving all the ${\cal K}_L$'s) which are due to
the second term in (\ref{eq:3.10}) actually sum up to give zero. Indeed we separate the d'Alembertian into a Laplacian and a second time-derivative, $\Box(\tilde r^B {\overline h})=\Delta(\tilde r^B {\overline h})- \tilde r^B\partial^2_u{\overline h}/c^2$, and we integrate by parts
the Laplacian using $\Delta y_L=l(l-1)\delta_{(i_l i_{l-1}}y_{L-2)}$,
generating in this way a Kronecker symbol (the surface term during the
integration by parts is zero by analytic continuation).  Since the $l$
indices $i_1...i_l$ are contracted with the $l$ indices carried by the
spatial gradients $\partial_L\equiv\partial_{i_1}...\partial_{i_l}$
present in the multipole expansion [see (\ref{eq:3.4})], the latter Kronecker symbol $\delta_{i_l i_{l-1}}$ generates a Laplacian
which is then equivalent (because it acts on a spherical retarded wave)
to a second time-derivative.  It is not difficult to check that the sum of all these terms with second-time-derivatives cancels exactly the other second-time-derivatives issued from the separation made above of the
d'Alembertian into a Laplacian.  Therefore we can re-write the multipole
decomposition by ignoring the second term in (\ref{eq:3.10}) and taking into
account only the first term which represents really the multipole moment
as generated by the (post-Newtonian expansion of the) source. Denoting by ${\cal H}_L$ the first term in (\ref{eq:3.10}) we obtain the main result of this paper (restoring the space-time indices $\mu\nu$):

\begin{equation}    
{\cal M}(h^{\mu\nu}) = {\rm FP}_{B=0}\, \Box^{-1}_R [
\tilde r^B {\cal M}(\Lambda^{\mu\nu})] - {4G\over c^4} \sum^{+\infty}_{l=0}
 {(-)^l\over l!} \partial_L \left\{ {1\over r} {\cal
 H}^{\mu\nu}_L (t-r/c) \right\} \ ,   \label{eq:3.11}
\end{equation} 
where the (source) multipole moments are given by the simple expression

\begin{equation}
 {\cal H}^{\mu\nu}_L (u) = {\rm FP}_{B=0}  
\int d^3 {\bf y} |\tilde {\bf y}|^B y_L \,
{\overline \tau}^{\mu\nu}({\bf y}, u)\ ,   \label{eq:3.12}
\end{equation} 
with $\overline \tau^{\mu\nu}$ the post-Newtonian expansion of
$\tau^{\mu\nu}$. In this expression there is no explicit $B$-factor
left out, so in contrast to (\ref{eq:3.10}) the integration over the
whole three-dimensional space contributes to the multipole moment,
including the regions at infinity, as well as the intermediate regions
and most importantly the near zone. As we shall see in Section IV, 
the first term in (\ref{eq:3.11}) represents the non-linear 
corrections to be added in a post-Minkowskian expansion to the (linear-looking) multipole expansion as given by the second term. Due to the presence of this first term the radiative multipole moments defined in the wave zone will differ from the source moments (see Section VI). It can be shown that the multipole expansion ${\cal M} (h^{\mu\nu})$ is actually independent of the length scale $r_0$ entering in $\tilde r =r/r_0$ and $\tilde {\bf y} = {\bf y}/r_0$ [namely, the $r_0$'s cancel out between the two terms of (3.11)].

Let us emphasize the interesting role played by the analytic continuation
all over the proof of (\ref{eq:3.12}). Witness in particular the
crucial passage from (\ref{eq:3.5}) to (\ref{eq:3.6}), which permits
{\it in fine} to get rid of the reference to the multipole expansion in
the integrand of the multipole moments themselves.  See also the last
step from (\ref{eq:3.10}) to (\ref{eq:3.12}) where we discard some
surface terms which are zero by analytic continuation, and which permits
to express the result in terms of an integral over the sole $\overline
\tau$ (without explicit reference to $\overline h$). The result (\ref{eq:3.11})-(\ref{eq:3.12}) generalizes (and clarifies) the result obtained by another method in paper II when
computing the multipole expansion of the gravitational field at the 2PN
order. In the Appendix A we present an alternative proof of the result (\ref{eq:3.11})-(\ref{eq:3.12}) based on the method of paper II.  

The multipole expansion (\ref{eq:3.11})-(\ref{eq:3.12}) has been written
using non trace-free multipole moments. Actually it is better to re-write it using symmetric and trace-free (STF) moments, because the non trace-free moments are not uniquely defined (for instance ${\cal K}_L$ and ${\cal H}_L$ yield the same multipole expansion). Here we simply report the result of the STF multipole expansion (which readily follows from the
equation (B.14a) in \cite{BD89}):

\begin{equation}
{\cal M}(h^{\mu\nu}) = {\rm FP}_{B=0}\, \Box^{-1}_R [
\tilde r^B {\cal M}(\Lambda^{\mu\nu})] - {4G\over c^4} \sum^{+\infty}_{l=0}
 {(-)^l\over l!} \partial_L \left\{ {1\over r} {\cal
 F}^{\mu\nu}_L (t-r/c) \right\} \ ,   \label{eq:3.13}
\end{equation} 
where the STF multipole moments are given by

\begin{equation}
{\cal F}^{\mu\nu}_L (u) = {\rm FP}_{B=0} \int d^3 {\bf y} |\tilde{\bf y}|^B
 {\hat y}_L \, \int^1_{-1} dz \delta_l(z) {\overline \tau^{\mu\nu}} 
 ({\bf y}, u+z|{\bf y}|/c)   \label{eq:3.14}
\end{equation} 
(${\hat y}_L$ denotes the trace-free part of $y_L$ \cite{N}). The STF
moments (\ref{eq:3.14}) involve an integration over the $z$-dependent
cone $t=u+z|{\bf y}|/c$ with weighting function

\begin{equation}
  \delta_l (z) = {(2l+1)!!\over 2^{l+1} l!} (1-z^2)^l
\ ; \quad\int^1_{-1} dz\delta_l (z) = 1\ ;
\quad {\rm lim}_{~l \to \infty}\delta_l (z)=\delta (z)\ . \label{eq:3.15}
\end{equation} 
As a check of the result let us consider the limit of linearized gravity
where we can neglect $\Lambda^{\mu\nu}$ and ${\cal M}(\Lambda^{\mu\nu})$,
and where $\tau^{\mu\nu}$ reduces to the matter stress-energy tensor
$T^{\mu\nu}$ with compact support, i.e.  $T^{\mu\nu}({\bf x},t)=0$ when
$|{\bf x}|>d$. In this limiting case the first term in (\ref{eq:3.13})
vanishes.  Furthermore replacing $\overline\tau^{\mu\nu}$ by $\overline
T^{\mu\nu}$ in (\ref{eq:3.14}) we can remove the factor $|\tilde{\bf y}|^B$
and the finite part at $B=0$ for compact-supported integrals.  Finally,
we can replace ${\overline T}^{\mu\nu}$ by $T^{\mu\nu}$ within the
compact support of a slowly-moving source (since $d\ll \lambda)$.  Thus
we recover exactly the expression of the multipole moments derived for
compact-support sources in the appendix B of \cite{BD89} and used for
studying the linearized gravity in \cite{DI91b} (in this case the
result is valid also for fast moving sources).

The multipole decomposition (\ref{eq:3.11})-(\ref{eq:3.15}) appears to be
quite general.  Physically it should apply to any isolated slowly-moving
source without singularities.  Technically it does not make any
reference, for instance, to the post-Minkowskian (or rather MPM)
expansion which has been invoked in order to derive it.  In a sense
(\ref{eq:3.11})-(\ref{eq:3.15}) represents a ``complete'' matching
equation (valid to any post-Newtonian and/or post-Minkowskian order),
which is the general consequence of our matching assumption (iv).  We
shall leave open the possibility that the multipole expansion
(\ref{eq:3.11})-(\ref{eq:3.15}), because of its generality, may have in
fact a domain of validity larger than the one of MPM approximate solutions.
For instance it is plausible that (\ref{eq:3.11})-(\ref{eq:3.15}) could
be proved in a more general context of exact solutions (using perhaps an
analysis similar to that of \cite{DS90}).

\section{The linearized multipolar metric} \label{sec:4}

The most general solution of the vacuum field equations (outside the
time-axis) was constructed within the MPM framework (see Section II).
Therefore, if the present analysis makes sense, it should be possible to
recast the general multipole expansion ${\cal M}(h^{\mu\nu} )$ as given
by (\ref{eq:3.13})-(\ref{eq:3.14}) into a form which shows clearly that it belongs to the class of MPM metrics. Namely, we would like to find a certain
``{\it linearized}~'' multipolar metric such that ${\cal M}(h^{\mu\nu})$
appears to be the post-Minkowskian iteration of that metric (in the MPM
sense).  The advantage is that the multipole moments parametrizing this
linearized metric, which constitute efficient tools in practical
computations of gravitational radiation (see e.g.
\cite{BDI95,B96,BIJ98}), will then be obtained with full generality as
computable functionals of the matter fields in the source.  This is what
we shall do in this section and the following one, where the linearized
multipolar metric associated with ${\cal M}(h^{\mu\nu})$ will be denoted
$h^{\mu\nu}_{\rm part.1}$ following the notation of paper I.

Let us denote the first term in (\ref{eq:3.13}) by

\begin{equation}
u^{\mu\nu} \equiv {\rm FP}_{B=0}\, \Box^{-1}_R [ \tilde r^B
{\cal M}(\Lambda^{\mu\nu})] \ .   \label{eq:4.1}
\end{equation} 
The first step is to compute the divergence $\partial_\nu u^{\mu\nu}$
of this term. To do this one notices first that ${\cal M}(h^{\mu\nu})$ is divergenceless by (\ref{eq:2.14}), and therefore that, by (\ref{eq:3.13}),

\begin{equation}
\partial_\nu u^{\mu\nu} = {4G\over c^4} \partial_\nu
\left( \sum^{+\infty}_{l =0} {(-)^l \over l !} \partial_L
\left\{ {1\over r} {\cal F}^{\mu\nu}_L (t-r/c) \right\} \right)\ .
\label{eq:4.2}
\end{equation} 
If analytic continuation factors $|\tilde {\bf y}|^B$ were absent in the
expression of ${\cal F}^{\mu\nu}_L$ given by (\ref{eq:3.14}), the
right-hand-side of (\ref{eq:4.2}) would be (formally) zero by virtue of
$\partial_\nu {\overline \tau}^{\mu\nu} = 0$.  With factors $|\tilde{\bf
y}|^B$ included, it is straightforward to find what is (\ref{eq:4.2}).  In fact the computation is the same as the one yielding the equation (4.1) in paper II.  One must evaluate the time derivative of ${\cal F}^{0\mu}_L$ using $\partial_\nu {\overline \tau}^{\mu\nu} = 0$, and perform some integrations by parts both with respect to ${\bf y}$ and $z$. We differentiate where appropriate the factor $|\tilde{\bf y}|^B$.  During the integrations by parts all the surface terms are zero by analytic continuation. Derivatives of the function $\delta_l (z)$ are computed using

\begin{mathletters}
\label{eq:4.3} \begin{eqnarray}
 {d\over dz} [\delta_{l +1} (z)] &=& -(2l +3) z\, \delta_l (z)
  \label{eq:4.3a}\\ {d^2\over dz^2} [\delta_{l +1} (z)] &=& -(2l
 +3)(2l +1)
  [\delta_l (z) - \delta_{l-1} (z)] \ . \label{eq:4.3b}
\end{eqnarray}
\end{mathletters} 
As a result, we find 

\begin{equation}
 \left( {d\over cdu} \right) {\cal F}^{0\mu}_L = l\, {\cal F} ^{\mu
 <i_l}_{L-1>} + {1\over 2l +3} \left( {d\over cdu} \right) ^2
 {\cal F}^{j\mu}_{jL} + {\cal G}^\mu_L \ , \label{eq:4.4}
\end{equation} 
where ${\cal F}^{\mu <i_l}_{L-1>}$ denotes the STF part of ${\cal F}^{\mu i_l}_{L-1}$, and where the function ${\cal G}^\mu_L$ is given by 

\begin{equation}
 {\cal G}^{\mu}_L (u)={\rm FP}_{B=0}\int d^3{\bf y} B |\tilde {\bf y}|^B |{\bf y}|^{-2}
 y_i \hat y_L \int^1_{-1} dz\,\delta_l (z) {\overline \tau}^{\mu i}
  ({\bf y}, u+z |{\bf y}|/c)\ . \label{eq:4.5}
\end{equation} 
With (\ref{eq:4.4}) in
hands, it is straightforward to transform (\ref{eq:4.2}) into

\begin{equation}
 \partial_\nu u^{\mu\nu} = {4G\over c^4} \sum^{+\infty}_{l =0}
{(-)^l \over l !} \partial_L \left\{ {1\over r}\, {\cal G}^{\mu}_L
(t-r/c) \right\} \ . \label{eq:4.6}
\end{equation} 
Another way of proving (\ref{eq:4.5})-(\ref{eq:4.6}) notices that the multipole expansion ${\cal M}(\Lambda^{\mu\nu})
\equiv \Lambda^{\mu\nu}_{\rm ext}$ is divergenceless by (\ref{eq:2.5}), and so from (\ref{eq:4.1}) we have $\partial_\nu u^{\mu\nu} = {\rm FP}_{B=0}\, \Box^{-1}_R [ B \tilde r^B r^{-1} n_i {\cal M} (\Lambda^{\mu i})]$,
where the factor $B$ comes from the derivation of $\tilde r^B$.
Thanks to this factor the finite part is actually a
residue, and we can perform an analysis analogous to the one 
going from (\ref{eq:3.2}) to (\ref{eq:3.9}) in the previous section. In this way we recover exactly the result (\ref{eq:4.5})-(\ref{eq:4.6}).

Having obtained the divergence of $u^{\mu\nu}$ in the form (\ref{eq:4.6}),
we proceed similarly to paper I and construct from (\ref{eq:4.6}) a
different object $v^{\mu\nu}$ which is like (\ref{eq:4.6}) a retarded
solution of the wave equation, and furthermore which is such that its
divergence cancels out the divergence of $u^{\mu\nu}$:  $\partial_\nu
v^{\mu\nu}=-\partial_\nu u^{\mu\nu}$.  In Appendix B we recall from
paper I (and \cite{B97quad}) the expression of
$v^{\mu\nu}$ in terms of STF tensors $A_L$, $B_L$, $C_L$ and $D_L$, and
we show the equivalence with the different expression:

\begin{mathletters} \label{eq:4.7}
\begin{eqnarray}
 v^{00} &=&{4G\over c^4}\left\{-{c\over r}\hbox{$\int$}{\cal
 G}^0+\partial_a \left( {1\over r} \left[ c \hbox{$\int$} {\cal G}^0_a
  + c^2 \hbox{$\int\!\!\int$} {\cal G}^a - {\cal G}^b_{ab} \right]
  \right) \right\}\ , \label{eq:4.7a}\\ 
v^{0i} &=& {4G\over c^4}
\left\{ - {1\over r} \left[ c\hbox{$\int$}
  {\cal G}^i - {1\over c} \dot{\cal G}^a_{ai} \right] + {c\over 2}
  \partial_a \left( {1\over r} \left[ \hbox{$\int$} {\cal G}^i_a -
  \hbox{$\int$} {\cal G}^a_i \right] \right)\right.  \nonumber\\
 &&\qquad   - \sum_{l \geq 2} {(-)^l\over l !} \partial_{L-1}
  \left( {1\over r} {\cal G}^0_{iL-1} \right) \biggr\} \ ,
  \label{eq:4.7b}\\ 
v^{ij} &=& {4G\over c^4} \biggl\{ {1\over r} {\cal G}^{(i}_{j)} 
  + 2 \sum_{l \geq 3} {(-)^l\over l !}\partial_{L-3}
  \left( {1\over rc^2} \ddot{\cal G}^a_{ijaL-3} \right)
  \nonumber\\ 
&& + \sum_{l \geq 2} {(-)^l\over l !} \left[ \partial_{L-2}
  \left( {1\over rc} \dot{\cal G}^0_{ijL-2} \right)
   +\partial_{aL-2} \left( {1\over r} {\cal G}^a_{ijL-2} \right)
   +2\delta_{ij}\partial_{L-1} \left( {1\over r} {\cal G}^a_{aL-1}
  \right) \right.  \nonumber\\ && \qquad  \left. \left.
   -4 \partial_{L-2(i} \left( {1\over r} {\cal G}^a_{j)aL-2} \right)
   -2\partial_{L-1}
    \left( {1\over r} {\cal G}^{(i}_{j)L-1} \right) \right] \right\}
    \ , \label{eq:4.7c} 
\end{eqnarray}
\end{mathletters}
where the ${\cal G}^\mu_L$'s are given by (4.5) (they
are all evaluated at the retarded time $u=t-r/c$).  The notation is
$T_{(ij)}={1\over 2} \left( T_{ij} + T_{ji} \right)$ and $T^{(i}_{j)}=
{1\over 2} \left( T_j^i + T_i^j \right)$;  ${\cal G}^\mu =({\cal G}^0,
{\cal G}^i)$ means ${\cal G}^\mu_L$ with $l =0$;  time derivatives are denoted by $\dot {\cal G} (u) = d{\cal G} (u)/du$, and time antiderivatives by $\int {\cal G}(u) = \int^u_{-\infty} dv {\cal G} (v)$, $\int\!\!\int {\cal G}(u) =\int^u_{-\infty} dv \int {\cal G}(v)$.  The main property of
$v^{\mu\nu}$, i.e.  $\partial_\nu v^{\mu\nu} =-\partial_\nu u^{\mu\nu}$,
is checked directly on (\ref{eq:4.7}).

With the latter construction of $v^{\mu\nu}$ we are able to define what
will constitute a ``linearized'' approximation to the multipolar
expansion ${\cal M}(h^{\mu\nu})$ given by (\ref{eq:3.13}).
Let us pose

\begin{equation}
 G h^{\mu\nu}_{\rm part.1} = -{4G\over c^4}
  \sum^{+\infty}_{l =0} {(-)^l \over l !} \partial_L \left\{
 {1\over r}{\cal F}^{\mu\nu}_L (t-r/c) \right\}-
 v^{\mu\nu}\ .\label{eq:4.8}
\end{equation} 
By the structure of $h^{\mu\nu}_{\rm part.1}$ made out of retarded
solutions of the source-free wave equation, and by the construction of
$v^{\mu\nu}$ ensuring its zero divergency [see (4.2)], we have (in $I\!\!R^3_* \times I\!\!R$)

\begin{eqnarray}
 \Box h^{\mu\nu}_{\rm part.1} &=& 0\ , \label{eq:4.9} \\
  \partial_\nu  h^{\mu\nu}_{\rm part.1} &=& 0\ . \label{eq:4.10}
\end{eqnarray} 
This means that (\ref{eq:4.8}) satisfies the linearized field equations
in the exterior region, i.e. (2.8)-(2.9) with $n=1$. By Theorem 2.1
of paper I [see also (4.7) in paper I], we know that the most general
solution of the system of equations (\ref{eq:4.9})-(\ref{eq:4.10}) can
always be written as the sum of a ``canonical'' linearized metric
$h^{\mu\nu}_{\rm can.1}$ (introduced by Thorne \cite{Th80}) and a
linear {\it gauge} transformation.  Therefore $h^{\mu\nu}_{\rm part.1}$
[which is referred to in paper I as a ``particular'' metric but which is
in fact quite general] reads

\begin{equation}
 h^{\mu\nu}_{\rm part.1} = h^{\mu\nu}_{\rm can.1} +
 \partial^\mu\varphi^\nu_1 + \partial^\nu\varphi^\mu_1 - \eta^{\mu\nu}
  \partial_\lambda\varphi^\lambda_1\ .\label{eq:4.11}
\end{equation}
The canonical metric is parametrized by {\it two} types of (STF) multipole
moments $I_L, J_L$,

\begin{mathletters} \label{eq:4.12}
\begin{eqnarray}
 h^{00}_{\rm can.1} &=& -{4\over c^2}\sum_{l\geq 0}
 {(-)^l\over l !} \partial_L \left( {1\over r} I_L (u)\right)\ ,
 \label{eq:4.12a}\\ h^{0i}_{\rm can.1} &=& {4\over c^3}\sum_{l\geq 1}
 {(-)^l\over l !} \left\{ \partial_{L-1} \left( {1\over r} \dot
 I_{iL-1} (u)\right) + {l\over l+1} \varepsilon_{iab}
 \partial_{aL-1} \left( {1\over r}
  J_{bL-1} (u)\right)\right\}\ , \label{eq:4.12b}\\ 
h^{ij}_{\rm can.1}
 &=&-{4\over c^4}\sum_{l\geq 2} {(-)^l\over l !} \left\{
 \partial_{L-2} \left( {1\over r}\ddot I_{ijL-2} (u)\right) +
 {2l\over l+1} \partial_{aL-2} \left( {1\over r}
 \varepsilon_{ab(i} \dot J_{j)bL-2}
 (u)\right)\right\}\ .\label{eq:4.12c} 
\end{eqnarray}
\end{mathletters}
On the other hand the vector associated with the gauge
transformation depends on {\it four} STF moments $W_L$, $X_L$,
$Y_L$, $Z_L$:

\begin{mathletters} \label{eq:4.13}
\begin{eqnarray}
 \varphi^0_1 &=& {4\over c^3}\sum_{l\geq 0} {(-)^l\over l !}
 \partial_L \left( {1\over r} W_L (u)\right)\ , \label{eq:4.13a}\\
 \varphi^i_1 &=& -{4\over c^4}\sum_{l\geq 0} {(-)^l\over l !}
 \partial_{iL} \left( {1\over r} X_L (u)\right) \nonumber \\ &&-{4\over
 c^4}\sum_{l\geq 1} {(-)^l\over l !} \left\{ \partial_{L-1}
 \left( {1\over r} Y_{iL-1} (u)\right) + {l\over l+1}
 \varepsilon_{iab} \partial_{aL-1}
  \left( {1\over r} Z_{bL-1} (u)\right)\right\}\ . \label{eq:4.13b}
\end{eqnarray}
\end{mathletters} 
All these multipole moments $I_L$, $J_L$, $W_L$, $X_L$, $Y_L$, $Z_L$ will be
computed in Section V.  Using the previous definition of a linearized metric
(\ref{eq:4.8})-(\ref{eq:4.13}), we can thus re-write our general
multipole expansion (\ref{eq:3.13}) as

\begin{equation}
 {\cal M} (h^{\mu\nu}) = Gh^{\mu\nu}_{\rm part.1} + u^{\mu\nu} +
 v^{\mu\nu} \  \label{eq:4.14}
\end{equation}
[where we recall that $u^{\mu\nu}$ and $v^{\mu\nu}$ are functionals given previously of ${\cal M}(\Lambda^{\mu\nu})$]. 
Intuitively from this equation, the terms $u^{\mu\nu}$ and $v^{\mu\nu}$
should represent the non-linear contributions (of order $G^2$ at least)
to be added to the ``linearized'' metric $h^{\mu\nu}_{\rm part.1}$ in
order to obtain the ``complete'' (including all powers of $G$)
multipole expansion ${\cal M} (h^{\mu\nu})$. We shall prove that this is
true, namely that by performing the non-linear iteration of
$h^{\mu\nu}_{\rm part.1}$ following exactly the MPM algorithm of paper I
[the only difference being that we have slightly modified the algorithm for the computation of $v^{\mu\nu}$, see Appendix B], we get an infinite power series in $G$ which agrees with ${\cal M} (h^{\mu\nu})$ term by term in the post-Minkowskian expansion.

Consistently with the algorithm of paper I we must consider that the 
first term in the post-Minkowskian expansion, i.e.  $Gh_{\rm part.1}$,
is purely of order $G$, and thus that
$h_{\rm part.1}$ itself is of zeroth order in $G$.  With this convention
let us show that the terms $u$ and $v$ in (\ref{eq:4.14}) are of order
$G^2$.  Suppose they are not, so that $u=Gu_{\rm part.1}+O(G^2)$ and
$v=Gv_{\rm part.1}+O(G^2)$ for some ``linearized'' coefficients $u_{\rm
part.1}$ and $v_{\rm part.1}$.  From (\ref{eq:4.14}) the multipole
expansion of the source, namely ${\cal M}(\Lambda) \equiv \Lambda({\cal
M}(h))$, is given by $\Lambda(G h_{\rm part.1}+u+v)$, and therefore,
inserting the previous assumptions for $u$ and $v$, by $\Lambda(G[h_{\rm
part.1}+u_{\rm part.1}+v_{\rm part.1}]+O(G^2))$.  Remember that
$\Lambda$ is quadratic in $h$, and pose $\Lambda(h)=N_2(h)+O(h^3)$
[see (2.8)-(2.9)].  Thus, obviously, $\Lambda({\cal M}(h))= G^2N_2(h_{\rm part.1}+u_{\rm part.1}+v_{\rm part.1})+O(G^3)$. Now, according to (\ref{eq:4.1}), the finite part of the retarded integral of
the source $\Lambda({\cal M}(h))$ is $u$ itself.  Using this fact and
the fact that the operator ${\rm FP} \Box^{-1}_R$ does not depend on $G$
(thus it does not mix up the powers of $G$), we obtain the equation
$u=G^2{\rm FP} \Box^{-1}_R [N_2(h_{\rm part.1}+u_{\rm part.1}+v_{\rm
part.1})]+O(G^3)$.  The right side of the latter equation is of order
$G^2$ so we deduce $u_{\rm part.1}=0$.  Then from the ``linearization''
of the formulas (\ref{eq:4.7}) we further deduce $v_{\rm part.1}=0$.
Thus $u$ and $v$ are indeed of order $O(G^2)$, and thus (\ref{eq:4.14}) shows that ${\cal M} (h^{\mu\nu})$ agrees with the ``particular'' metric at
linearized order:

\begin{equation}
 {\cal M} (h^{\mu\nu}) = Gh^{\mu\nu}_{\rm part.1} + O(G^2)\ .
 \label{eq:4.15}
\end{equation} 
Now let us denote $u=G^2u_{\rm part.2}+O(G^3)$ and $v=G^2v_{\rm
part.2}+O(G^3)$.  By the equation used just before (4.15),
in which we can now insert
$u_{\rm part.1}=v_{\rm part.1}=0$, we have $u=G^2{\rm FP} \Box^{-1}_R
[N_2(h_{\rm part.1})] +O(G^3)$ thus

\begin{equation}
u^{\mu\nu}_{\rm part.2} = {\rm FP}_{B=0}\,
\Box^{-1}_R [\tilde r^B N_2^{\mu\nu}(h_{\rm part.1})] \ .   \label{eq:4.16}
\end{equation} 
Next $v_{\rm part.2}$ is obtained from $u_{\rm part.2}$ by the formulas
(\ref{eq:4.7}) [see also Appendix B], which are precisely those used in
the algorithm of paper I (as redefined in \cite{B97quad}).  The
definition of the quadratic part of the particular metric in paper I was
$h_{\rm part.2}=u_{\rm part.2}+v_{\rm part.2}$ [indeed see (\ref{eq:2.10})--(\ref{eq:2.12})],
so we find from (\ref{eq:4.14}) that ${\cal M} (h^{\mu\nu})$ agrees with
the particular metric to quadratic order,

\begin{equation}
 {\cal M} (h^{\mu\nu}) = Gh^{\mu\nu}_{\rm part.1} + G^2h^{\mu\nu}_{\rm
 part.2} + O(G^3)\ . \label{eq:4.17}
\end{equation} 
The same reasoning is easily extended to all orders in $G$. 

In conclusion the ``particular'' metric $h^{\mu\nu}_{\rm
part}=Gh^{\mu\nu}_{\rm part.1}+G^2h^{\mu\nu}_{\rm part.2}+O(G^3)$ which
was defined in paper I agrees, in the sense of powers series in
$G$, with the general multipole expansion ${\cal M} (h^{\mu\nu})$.  This
result is mandatory because it was shown in paper I (Theorem 4.2)
that the general solution of the vacuum Einstein equations in
$I\!\!R^3_* \times I\!\!R$ can be written as $h^{\mu\nu}_{\rm
part}$ for {\it some} set of moments $I_L$, $J_L$, $W_L$, $X_L$, $Y_L$,
and $Z_L$.  This is of course consistent with the fact that we have made
no restriction when deriving the multipole expansion ${\cal M}(h^{\mu\nu})$
except that it should correspond to a slowly-moving isolated system
(without singularities).  What we have gained with respect to paper I is
that we understand from Section III the relation between ${\cal M}
(h^{\mu\nu}) \equiv h^{\mu\nu}_{\rm part}[I,J,...,Z]$ and the matter
distribution in the source, and therefore that we are able to {\it
compute} the multipole moments $I_L$, $J_L$, ..., $Z_L$ (given a post-Newtonian algorithm for the computation of ${\overline \tau}^{\mu\nu}$).

\section{The irreducible multipole moments}     
\label{sec:5}

By (\ref{eq:4.11})-(\ref{eq:4.13}) the linearized metric
$h^{\mu\nu}_{\rm part.1}$ is parametrized by six sets of irreducible
(STF) multipole moments, with two sets $I_L$, $J_L$ parametrizing the
``canonical'' linear metric $h^{\mu\nu}_{\rm can.1}$, and four sets
$W_L$, $X_L$, $Y_L$, $Z_L$ parametrizing a gauge transformation.  We
can refer to the moments $I_L$, $J_L$, $W_L$, $X_L$,
$Y_L$, $Z_L$ as the source multipole moments.  Of course since the
moments $W_L$, $X_L$, $Y_L$, $Z_L$ parametrize a gauge transformation,
they do not play a physical role at the level of the {\it linearized}
approximation.  In this sense, these four moments are ``less important''
than the moments $I_L$ and $J_L$, which constitute the ``main''
multipole moments of the source, respectively of mass-type and
current-type. However, it is important to keep the moments $W_L,\cdots,
Z_L$ as they start playing a role at the nonlinear level (at a high
post-Newtonian approximation \cite{B96}). See the discussion in Section VI
where we recall also that the six sets of moments $I_L$, $J_L$, $W_L,...,Z_L$
are in fact equivalent physically to only two sets of other moments
$M_L$ and $S_L$.

In the present section we use the results of Section III to compute
explicitly the moments $I_L$, $J_L,\cdots,Z_L$.  For this
purpose it suffices to decompose into irreducible pieces the functions
${\cal F}^{\mu\nu}_L$ and ${\cal G}^\mu_L$ which parametrize the
multipole expansions (\ref{eq:3.13}) and (\ref{eq:4.2}).  We first
decompose the components of ${\cal F}^{\mu\nu}_L$ according to

\begin{mathletters} \label{eq:5.1}
\begin{eqnarray}
  {\cal F}^{00}_L &=& R_L \ , \label{eq:5.1a} \\ {\cal F}^{0i}_L &=&
  ^{(+)}T_{iL} + \varepsilon_{ai<i_l} {}^{(0)}T_{L-1>a}
  +\delta_{i<i_l} {}^{(-)}T_{L-1>}\ ,\label{eq:5.1b}\\ {\cal
  F}^{ij}_L &=& ^{(+2)}U_{ijL} + \displaystyle{\mathop{STF}_L}\
   \displaystyle{\mathop{STF}_{ij}}\, [\varepsilon_{aii_l}
  {}^{(+1)}U_{ajL-1} +\delta_{ii_l} {}^{(0)}U_{jL-1}\nonumber \\ &&
  +\delta_{ii_l} \varepsilon_{aji_{l-1}} {}^{(-1)}U_{aL-2}
  +\delta_{ii_l} \delta_{ji_{l-1}} {}^{(-2)}U_{L-2} ] +
  \delta_{ij} V_L \ , \label{eq:5.1c} \end{eqnarray}
\end{mathletters}
where the ten tensors $R_L$, $^{(+)}T_{L+1}$,\dots, $^{(-2)}U_{L-2}$,
$V_L$ are STF in all their indices. We use a notation standard from the
works \cite{BD86,DI91b} (notably $<>$ denotes the STF projection \cite{N}).
These ten tensors are uniquely given in terms of the ${\cal
F}^{\mu\nu}_L$'s by the inverse formulas

\begin{mathletters} \label{eq:5.2}
\begin{eqnarray}
 R_L&=& {\cal F}^{00}_L \ , \label{eq:5.2a}\\ ^{(+)}T_{L+1}&=& {\cal
 F}^{0<i_{l+1}}_{L>} \ , \label{eq:5.2b}\\ ^{(0)}T_L&=& {l \over
 l +1}
  {\cal F}^{0a}_{b<L-1}\, \varepsilon_{i_l >ab} \ ,
  \label{eq:5.2c}\\ ^{(-)}T_{L-1}&=& {2l-1 \over 2l +1}
  {\cal F}^{0a}_{aL-1}\ , \label{eq:5.2d}\\ ^{(+2)}U_{L+2}&=& {\cal
 F}^{<i_{l+2}i_{l +1}}_{L>}\ ,\label{eq:5.2e}\\ ^{(+1)}U_{L+1}&=&
 {2l \over l +2}\ \displaystyle{\mathop{STF}_{L+1}} \  {\cal
 F}^{<ci_l>}_{dL-1}\varepsilon_{i_{l+1}cd}\ , \label{eq:5.2f}\\
 ^{(0)}U_L&=& {6l(2l-1) \over (l+1)(2l+3)}\
 \displaystyle{\mathop{STF}_L}\ {\cal
 F}^{<ai_l>}_{aL-1}\ ,\label{eq:5.2g}\\ ^{(-1)}U_{L-1}&=&
 {2(l-1)(2l-1)\over (l+1)(2l+1)}
  \ \displaystyle{\mathop{STF}_{L-1}}\ {\cal
  F}^{<ac>}_{bcL-2}\varepsilon_{i_{l-1}ab}\ , \label{eq:5.2h}\\
 ^{(-2)}U_{L-2}&=& {2l-3 \over 2l+1}
  {\cal F}^{<ab>}_{abL-2}\ , \label{eq:5.2i}\\ V_L&=& {1\over 3}\,
  {\cal F}^{aa}_L\ . \label{eq:5.2j} \end{eqnarray}
\end{mathletters}
See for instance (5.5)-(5.8) in \cite{DI91b}.  Next we
decompose the tensors ${\cal G}^\mu_L$ according to

\begin{mathletters} \label{eq:5.3}
\begin{eqnarray}
 {\cal G}^0_L &=& P_L \ , \label{eq:5.3a} \\
{\cal G}^i_L &=&
 ^{(+)}Q_{iL} + \varepsilon_{ai<i_l} {}^{(0)}Q_{L-1>a}
    + \delta_{i<i_l}{}^{(-)}Q_{L-1>}\ , \label{eq:5.3b}
\end{eqnarray}
\end{mathletters} 
with inverse formulas

\begin{mathletters} \label{eq:5.4}
\begin{eqnarray}
 P_L&=& {\cal G}^{0}_L \ , \label{eq:5.4a}\\
 ^{(+)}Q_{L+1}&=& {\cal  G}^{<i_{l+1}}_{L>} \ , \label{eq:5.4b}\\
 ^{(0)}Q_L&=& {l \over l+1}
  {\cal G}^{a}_{b<L-1}\, \varepsilon_{i_l >ab} \ , \label{eq:5.4c}\\
 ^{(-)}Q_{L-1}&=& {2l-1 \over 2l +1}
  {\cal G}^{a}_{aL-1}\ . \label{eq:5.4d} \end{eqnarray}
\end{mathletters} 
The tensors parametrizing ${\cal G}^\mu_L$
are not independent of the tensors parametrizing ${\cal F}^{\mu\nu}_L$.
This is because the metric $h^{\mu\nu}_{\rm part.1}$ is divergenceless
by (\ref{eq:4.10}). The (four) relations linking these tensors are
readily obtained from (\ref{eq:4.4}). We have

\begin{mathletters} \label{eq:5.5} \begin{eqnarray}
 P_L &=& {1\over c}\,\dot R_L - l\,{}^{(+)} T_L - {1\over
 c^2(2l+1)}
  \,{}^{(-)}\ddot T_L\ , \label{eq:5.5a}\\ ^{(+)}Q_L &=& {1\over
 c}\,{}^{(+)}\dot T_L - (l-1){}^{(+2)} U_L
   - {(l+1)(2l+3)\over 6c^2l(2l-1)(2l+1)}
  \,{}^{(0)}\ddot U_L - {1\over c^2(2l+1)}\, \ddot V_L\ ,
 \label{eq:5.5b}\\ ^{(0)}Q_L &=& {1\over c}\,{}^{(0)}\dot T_L -
 {l\over 2}{}^{(+1)} U_L
   - {l+2\over 2c^2(l+1)(2l+1)}
     \,{}^{(-1)}\ddot U_L\ , \label{eq:5.5c}\\ ^{(-)}Q_L &=& {1\over
 c}\,{}^{(-)}\dot T_L - {l+1\over 6}
    {}^{(0)} U_L - {1 \over c^2(2l+3)}
     \,{}^{(-2)}\ddot U_L - (l+1) V_L \ .\label{eq:5.5d}
\end{eqnarray} \end{mathletters} 
These relations permit to express the ten independent components of
$h^{\mu\nu}_{\rm part.1}$ in terms of only six independent combinations
of STF tensors.  We subsitute the decompositions (\ref{eq:5.1}) and
(\ref{eq:5.3}) into the definition of the linearized metric
(\ref{eq:4.7})-(\ref{eq:4.8}).  After some manipulations of STF tensors,
and use of the previous relations (\ref{eq:5.5}), we arrive at a rather
messy expression but which can be compared directly with the general
decomposition of a linearized metric as
given by the equation (2.25) in paper I.  Then the six
sets of multipole moments $I_L$, $J_L$, $W_L$, $X_L$, $Y_L$, $Z_L$ entering $h^{\mu\nu}_{\rm part.1}$ are
obtained by applying the definitions (2.26) in paper I (actually our
definitions differ from paper I by some constant factors).  First the
moments $I_L$ and $J_L$ are obtained as follows.  In the particular
cases where $I_L$ has zero or one index ($l = 0,1$) and where $J_L$ has
one index ($l = 1$), we have

\begin{mathletters} \label{eq:5.6} \begin{eqnarray}
 I &=& {1\over c^2} (R +3V) - {4\over c^3}{}^{(-)}\dot T + {1\over c^4}
   {}^{(-2)}\ddot U - {1\over c} \hbox{$\int$} P + {3\over c^2}{}^{(-)}
   Q\ , \label{eq:5.6a} \\ I_i &=& {1\over c^2}(R_i +3V_i) -{2\over
 c^3}{}^{(-)}\dot T_i+{1\over 3c^4}
  {}^{(-2)}\ddot U_i - {1\over c} \hbox{$\int$} P_i-
  \hbox{$\int\!\!\int$} {}^{(+)}Q_i + {5\over 3c^2}{}^{(-)}
  Q_i\ ,\label{eq:5.6b} \\
   J_i &=& -{2\over c} {}^{(0)}T_i + {1\over 2c^2} {}^{(-1)}\dot U_i +
   2 \hbox{$\int$} {}^{(0)} Q_i\ . \label{eq:5.6c} \end{eqnarray}
\end{mathletters} 
Then, in the generic case where $I_L$ and $J_L$ have
at least two indices ($l \geq 2$), we have

\begin{mathletters} \label{eq:5.7} \begin{eqnarray}
 I_L &=& {1\over c^2} (R_L + 3V_L) - {4\over c^3(l +1)} {}^{(-)}
 \dot T_L+ {2\over c^4(l +1)(l+2)}{}^{(-2)} \ddot U_L
    \ , \label{eq:5.7a} \\ J_L &=& -{l+1\over l c}{} ^{(0)}T_L
 +{1\over 2l c^2}
     {}^{(-1)} \dot U_L \ . \label{eq:5.7b} \end{eqnarray}
\end{mathletters} 
Secondly, the four tensors $W_L$, $X_L$, $Y_L$, $Z_L$
are generically obtained as

\begin{mathletters} \label{eq:5.8} \begin{eqnarray}
  W_L &=& {1\over c(l+1)}{}^{(-)}T_L-{1\over 2c^2(l+1)(l+2)}
  {}^{(-2)}\dot U_L\ , \label{eq:5.8a}\\ X_L &=& {1\over
  2(l+1)(l+2)}{}^{(-2)} U_L\ ,\label{eq:5.8b}\\ Y_L &=& {3\over
  c(l+1)}{}^{(-)}\dot T_L-{2\over c^2(l+1)(l+2)}
  {}^{(-2)}\ddot U_L - 3\, V_L\ , \label{eq:5.8c}\\
   Z_L &=& -{1\over 2l}{}^{(-1)}U_L\ . \label{eq:5.8d}
\end{eqnarray}
\end{mathletters} 
The expressions of all these moments
are obtained first by substituting into their definitions
(\ref{eq:5.6})-(\ref{eq:5.8}) the inverse relations (\ref{eq:5.2}) and
(\ref{eq:5.4}), and second by using the expressions
(\ref{eq:3.14}) and (\ref{eq:4.5}) of the functions
${\cal F}^{\mu\nu}_L$ and ${\cal G}^{\mu}_L$.

We first investigate the lowest-order moments $I$, $I_i$ and $J_i$
defined by (\ref{eq:5.6}), which can respectively be called the mass 
monopole (or total ADM mass), mass dipole and current dipole 
of the source. As is
readily checked using (\ref{eq:5.5}) (and the fact that
${}^{(0)}U={}^{(+1)} U_i=0$), we have the conservation laws
appropriate for gravitational monopole and dipoles, i.e.

\begin{equation}
 \dot I = 0\ , \qquad \ddot I_i = 0\ , \qquad \dot J_i = 0\ .
 \label{eq:5.9}
\end{equation} 
For simplicity we analyse only the case of the mass monopole $I$;  the
analysis of the dipoles $I_i$ and $J_i$ is similar.  The expression of
$I$ deduced from (\ref{eq:5.6a}) is

\begin{eqnarray}
 Ic^2 &=& \hbox{FP}_{B=0} \text{$\int$} d^3{\bf y}|\tilde{\bf y}|^B
 \int^1_{-1} dz \Bigl\{ \delta_0 (\overline\tau^{00} 
+\overline\tau^{ii}) -{4\over 3c}
 \delta_1 y_i \dot{\overline\tau}^{0i} + {1\over 5c^2} 
\delta_2 \hat y_{ij}
 \ddot{\overline\tau}^{ij}
    \nonumber\\ &&  \qquad \qquad + B|{\bf y}|^{-2} (\delta_1 y_{ij}
 \overline\tau^{ij}
  - c\delta_0 y_i \hbox{$\int$}\overline\tau^{0i} ) \Bigr\}
   ({\bf y},u+z |{\bf y}|/c) \ , \label{eq:5.10}
\end{eqnarray}
which can be transformed using (\ref{eq:4.3}) into the simpler form 

\begin{equation}
 Ic^2 =\hbox{FP}_{B=0} \text{$\int$} d^3{\bf y}|\tilde{\bf y}|^B \int^1_{-1}
  dz \delta_0 \left\{\overline\tau^{00}
 - z|{\bf y}|^{-1} y_i\overline{\tau}^{0i} - Bc |{\bf
y}|^{-2} y_i \hbox{$\int$} \overline\tau^{0i} \right\}
  ({\bf y},u+z |{\bf y}|/c)\ . \label{eq:5.11}
\end{equation} 
The latter expression looks unfamiliar for a conserved mass, but this is
simply due to the unusual spacelike hypersurface $t-z|{\bf y}|/c
=u=$~const on which one integrates (see the discussion in \cite{DI91b}).
The following
technical identity is useful to transform (\ref{eq:5.11}):

\begin{equation}
 {d\over dz'} \{ ( \overline\tau^{00} -z'|{\bf y}|^{-1}
 y_i\overline\tau^{0i})({\bf
 y},u+z' |{\bf y}|/c)\} = - \partial_i \{ |{\bf y}| \overline\tau^{0i} 
 ({\bf  y},u+z'|{\bf y}| /c)\}\ .  \label{eq:5.12}
\end{equation} 
Multiplying this identity by $|\tilde{\bf y}|^B$, integrating over ${\bf y}$, and over $z'$ from 0 to $z$, and then multiplying by $\delta_0$ and integrating over $z$ from $-1$ to 1, permits to find the equivalent of (\ref{eq:5.11}) when one uses the usual spacelike hypersurface $t=u=$~const,

\begin{equation}
 Ic^2 = \hbox{FP}_{B=0} \text{$\int$} d^3{\bf y} |\tilde{\bf y}|^B \left[
  \overline\tau^{00} - Bc |{\bf y}|^{-2} y_i \hbox{$\int$}
\overline\tau^{0i} \right] ({\bf y},u)\ .  \label{eq:5.13}
\end{equation}
The fact that $I$ is constant is easily checked on this expression.
Note that the second term in (\ref{eq:5.13}) [whose time-derivative is
associated with the flux of radiation at infinity] involves a factor $B$
and therefore its value comes only from the poles of the integral at the
upper bound $|{\bf y}| \to +\infty$.

We now deal with the ``dynamic'' moments $I_L$ and $J_L$ having
$l\geq 2$. In order to express them, it is convenient to use the
following notation for combinations of components of the pseudo-tensor
$\overline\tau^{\mu\nu}$:  

\begin{mathletters} \label{eq:5.14}
\begin{eqnarray}
 \overline{\Sigma} &=& {\overline\tau^{00} +\overline\tau^{ii}\over c^2}\ ,
\label{eq:5.14a}\\
 \overline{\Sigma}_i &=& {\overline\tau^{0i}\over c}\ , \label{eq:5.14b}\\
\overline{\Sigma}_{ij} &=& \overline{\tau}^{ij}\ , \label{eq:5.14c}
 \end{eqnarray}
\end{mathletters} 
where $\overline{\tau}^{ii} \equiv\delta_{ij}\overline\tau^{ij}$.
($\overline{\Sigma}$, $\overline{\Sigma}_i$ and $\overline{\Sigma}_{ij}$
are of zeroth order in the post-Newtonian expansion.) Then, by
(\ref{eq:5.7}), (\ref{eq:5.2}) and (\ref{eq:3.14}), we find

\begin{eqnarray}
 I_L(u)&=& \hbox{FP}_{B=0} \text{$\int$} d^3{\bf y}|\tilde{\bf y}|^B
 \int^1_{-1} dz\left\{ \delta_l\hat y_L\overline\Sigma -{4(2l+1)\over
  c^2(l+1)(2l+3)} \delta_{l+1} \hat y_{iL} \dot{\overline\Sigma}_i\right.
  \nonumber\\
 &&\qquad\qquad \left. +{2(2l+1)\over
  c^4(l+1)(l+2)(2l+5)} \delta_{l+2} \hat y_{ijL}
  \ddot{\overline\Sigma}_{ij} \right\} ({\bf y},u+z |{\bf y}|/c)\ ,
  \label{eq:5.15}\\ \nonumber \\ 
J_L(u)&=& \varepsilon_{ab<i_l} \hbox{FP}_{B=0} \text{$\int$}
   d^3{\bf y}|\tilde{\bf y}|^B \int^1_{-1} dz\biggl\{ \delta_l\hat
  y_{L-1>a} \overline\Sigma_b   \nonumber\\ 
&&\qquad\qquad  -{2l+1\over c^2(l+2)(2l+3)} \delta_{l+1} \hat y_{L-1>ac}    \dot{\overline\Sigma}_{bc}
  \biggr\} ({\bf y},u+z |{\bf y}|/c)\ . \label{eq:5.16}
\end{eqnarray}
These expressions have been derived in the non-linear theory (to all orders
in the post-Newtonian expansion).  As a check of $I_L$ and $J_L$ we can
compare their expressions to the corresponding expressions derived by
Damour and Iyer \cite{DI91b} in the case of the linearized theory, where
we can replace the pseudo-tensor $\overline\tau^{\mu\nu}$ by the matter
stress-energy tensor $T^{\mu\nu}$ in flat space-time (we have $\overline
T^{\mu\nu}= T^{\mu\nu}$ inside the slowly-moving source), and then
remove the analytic continuation factors since $T^{\mu\nu}$ is
compact-supported.  We find perfect agreement with the equations (5.33)
and (5.35) in \cite{DI91b}.  In the non-linear theory but at the 2PN
order, the expressions (\ref{eq:5.15})-(\ref{eq:5.16}) were already
derived in paper II.  To the 1PN order these expressions are equivalent
to some different expressions derived earlier in \cite{BD89,DI91a} (see
paper II for the proof).

Finally we write down the four other moments
$W_L,\cdots,Z_L$. They are easily obtained as 

\begin{eqnarray}
 W_L(u)&=& \hbox{FP}_{B=0} \int d^3{\bf y}|\tilde{\bf y}|^B \int^1_{-1}
  dz\left\{ {2l+1\over (l+1)(2l+3)} \delta_{l+1} \hat
  y_{iL} \overline\Sigma_i\right. \nonumber\\ 
&&\qquad\qquad \left.
  -{2l+1\over 2c^2(l+1)(l+2)(2l+5)} \delta_{l+2} \hat y_{ijL} 
  \dot{\overline\Sigma}_{ij} \right\} ({\bf y},u+z |{\bf y}|/c)\ ,
  \label{eq:5.17}\\ \nonumber\\
 X_L(u)&=& \hbox{FP}_{B=0} \int
  d^3{\bf y}|\tilde{\bf y}|^B \int^1_{-1}
  dz\left\{  {2l+1\over 2(l+1)(l+2)(2l+5)} \delta_{l+2}
  \hat y_{ijL} \overline\Sigma_{ij} \right\} ({\bf y},u+z |{\bf y}|/c)\ ,
  \nonumber \\ \label{eq:5.18}\\ \nonumber \\
Y_L(u)&=& \hbox{FP}_{B=0} \int d^3{\bf y}|\tilde{\bf y}|^B \int^1_{-1}
  dz\left\{ -\delta_l \hat y_L \overline\Sigma_{ii} + {3(2l+1)\over
  (l+1)(2l+3)} \delta_{l+1} \hat y_{iL} \dot{\overline\Sigma}_i\right.
  \nonumber\\ 
&&\qquad\qquad \left.- {2(2l+1)\over
  c^2(l+1)(l+2)(2l+5)} \delta_{l+2} \hat y_{ijL}
  \ddot{\overline\Sigma}_{ij} \right\} ({\bf y},u+z |{\bf y}|/c)\ ,
  \label{eq:5.19}\\  \nonumber \\
 Z_L(u)&=& \hbox{FP}_{B=0} \int
  d^3{\bf y}|\tilde{\bf y}|^B \int^1_{-1}
  dz\left\{- {2l+1\over (l+2)(2l+3)} \varepsilon_{ab<i_l}
  \delta_{l+1} \hat y_{L-1>bc} \overline\Sigma_{ac} \right\} ({\bf y},u+z
  |{\bf y}|/c)\ . \nonumber \\
          \label{eq:5.20}
\end{eqnarray}
To Newtonian order these expressions are in agreement with the equations
(4.17) in \cite{B96}.

\section{Discussion} \label{sec:6}

Of what use are the expressions of the STF multipole moments
$I_L$, $J_L$ and $W_L$, $X_L$, $Y_L$, $Z_L$ obtained in the previous
equations (\ref{eq:5.15})-(\ref{eq:5.20})?  From
(\ref{eq:4.11})-(\ref{eq:4.13}) these moments parametrize the linearized
metric $h^{\mu\nu}_{\rm part.1}$ which is the ``seed'' of the infinite
nonlinear algorithm of paper I. Thus for a specific application the expressions (\ref{eq:5.15})-(\ref{eq:5.20}) have to be computed in a
post-Newtonian expansion up to a given order and for a specific matter
model (i.e.  a specific choice of $T^{\mu\nu}$), and then inserted into
the so-called ``particular'' algorithm of paper I for the computation of
the field non-linearities (essentially this is what has been done for
compact binary systems in \cite{BDI95,B96,BIJ98}).  Actually the main
moments to be computed are $I_L$ and $J_L$ because the other moments
$W_L,\cdots, Z_L$ parametrize a gauge transformation and thus have no
physical implications at the linearized order.  In terms of a
post-Newtonian expansion it was shown in \cite{B95,B96} that up to the
2PN order it is sufficient to compute $I_L$ and $J_L$, but that the
other moments $W_L,\cdots, Z_L$ start contributing at the 2.5PN order.

Now it was proved in paper I that the full nonlinear metric outside an
isolated system can always be parametrized (modulo a coordinate
transformation) by only {\it two} sets of STF multipole moments, say
$M_L$ and $S_L$ (different from $I_L$ and $J_L$).  These moments
parametrize the so-called ``canonical'' algorithm of paper I, defined by
the same formulas as for the ``particular'' algorithm but starting with
the canonical linearized metric $h^{\mu\nu}_{\rm can.1}$ given by
(\ref{eq:4.12}) (but where $I_L$, $J_L$ are replaced by $M_L$, $S_L$). By Theorems 4.2 and 4.5 in paper I the canonical and particular algorithms differ from each other by a coordinate transformation. Therefore the multipole moments $M_L$ and $S_L$ are necessarily given as some functionals of the other moments $I_L$, $J_L$ and $W_L,\cdots ,Z_L$, i.e.

\begin{mathletters} \label{eq:6.1}
\begin{eqnarray}
 M_L &=& M_L [ I, J, W, X, Y, Z]\ , \label{eq:6.1a}\\ 
 S_L &=& S_L [ I, J, W, X, Y, Z]\ . \label{eq:6.1b}
\end{eqnarray}
\end{mathletters}
These functionals are quite complicated in general but can
be explicitly constructed up to any post-Minkowskian order by
implementing the coordinate transformation between the two harmonic
coordinate systems in which the ``particular'' and ``canonical'' metrics
are defined (see Section~4.3 in paper I).  Since at the linearized level
the canonical and particular metrics differ by a mere gauge
transformation [see (\ref{eq:4.11})], we have agreement at
this level between $M_L$, $S_L$ and $I_L$, $J_L$:

\begin{mathletters} \label{eq:6.2}
\begin{eqnarray}
 M_L &=& I_L + O(G)\ , \label{eq:6.2a}\\ 
 S_L &=& J_L + O(G)\ , \label{eq:6.2b}
\end{eqnarray}
\end{mathletters} 
where $O(G)$ symbolizes some nonlinear (quadratic at least) products 
of the source moments. Furthermore, the result of \cite{B96} is that 
in a post-Newtonian re-expansion $c\to +\infty$ we have

\begin{equation}
 M_L = I_L + {1\over c^5}\delta I_L + O\left(1\over c^6\right)\ ,
\label{eq:6.3}
\end{equation}  
where $\delta I_L$ is given in details by the equation (4.24) of \cite{B96}.

On the other hand, it is known (see e.g.  \cite{Th80}) that the transverse
and tracefree (TT) part of the spatial metric at leading order $1/R$ in the distance can be parametrized by still another double set of STF multipole moments, say $U_L$ and $V_L$.  These moments are called the radiative moments of the source, as they are the moments which would be measured at infinity. The radiative moments differ from the source moments because of the non-linear terms $u^{\mu\nu}$ and $v^{\mu\nu}$ in (\ref{eq:4.14}).
Since the exterior field is entirely determined by the moments
$M_L$, $S_L$, the radiative moments $U_L$, $V_L$ are necessarily given
as some (fully non-linear) functionals of them:

\begin{mathletters} \label{eq:6.4}
\begin{eqnarray}
 U_L &=& U_L [M, S]\ , \label{eq:6.4a}\\ 
 V_L &=& V_L [M, S]\ . \label{eq:6.4b} 
\end{eqnarray}
\end{mathletters} 
$U_L$ and $V_L$ are conveniently chosen in such a way that at
the linearized order they reduce to the $l$th time derivatives of the
moments $M_L$, $S_L$:

\begin{mathletters} \label{eq:6.5}
\begin{eqnarray}
 U_L &=& {d^l M_L \over du^l} + O(G) \ , \label{eq:6.5a}\\ 
 V_L &=& {d^l S_L \over du^l} + O(G) \ , \label{eq:6.5b} 
\end{eqnarray}
\end{mathletters} 
where $O(G)$ denotes the nonlinear terms.  It was shown in \cite{B87}
that the functionals (\ref{eq:6.4}) can be constructed to all
orders in the post-Minkowskian expansion by implementing the coordinate
transformation between the harmonic coordinates and some suitable
``radiative'' coordinates in which the metric admits an expansion in
powers of $1/R$ (without logarithms of $R$).  Furthermore, once obtained
in a post-Minkowskian expansion, the functionals (\ref{eq:6.4}) can be
re-expanded when $c \to +\infty$.  In this limit the dominant correction
is of order $1/c^3$ (or, rather, $G/c^3$) and due to the so-called tails
of waves \cite{BD92}.  For instance we have, in the quadrupole case,
 
\begin{equation}
 U_{ij}(u) = {d^2M_{ij}\over du^2}(u)
+ {2GM\over c^3}\int^u_{-\infty} dv \left[\ln \left({u-v}\over {2b}\right)
+{11\over 12}\right] {d^4M_{ij}\over du^4}(v) + O\left(1\over c^5\right)\ ,
\label{eq:6.6}
\end{equation}  
where $M$ is the ADM mass of the source ($M\equiv I$), and where $b$ is
a constant time scale entering the relation between harmonic and
radiative coordinates.  The complete correction of order $G$ involves
other terms and can be found in \cite{BD92,B97quad}.

We now understand that the explicit expressions (\ref{eq:5.15})-(\ref{eq:5.20})
of the ``source'' multipole moments $I_L$, $J_L$ and $W_L$,...,$Z_L$ are
to be inserted into the chain of functionals (\ref{eq:6.1}) and
(\ref{eq:6.4}) giving the radiative moments $U_L$, $V_L$ detected far
away from the source.  Since by the present investigation the source
moments have been related to the stress-energy tensor of the source up
to any (post-Newtonian) order, and since the functionals (\ref{eq:6.1})
and (\ref{eq:6.4}) can be ``algorithmically'' computed up to any
nonlinear order \cite{BD86,B87} (and then be re-expanded when $c \to
\infty$), we conclude that the radiative moments $U_L$, $V_L$ can be
computed up to any post-Newtonian order (in principle), in terms of the
source parameters.  Of course the resulting formalism becomes extremely
complicated when going to high post-Newtonian orders, especially when
computing the source multipole moments (\ref{eq:5.15})-(\ref{eq:5.16}).
For the moment it has been implemented in the case of compact binary
systems up to the 3PN level only \cite{BDI95,B96,BIJ98}.

\acknowledgments

The author would like to thank Thibault Damour and Bernd Schmidt 
for reading the
manuscript, and for discussion and useful comments.

\appendix
\section{Alternative proof of the result}
 \label{sec:apa}

In this Appendix we present an alternative derivation of the main result of the paper, which is the multipole decomposition (\ref{eq:3.11})-(\ref{eq:3.12}) or equivalently (\ref{eq:3.13})-(\ref{eq:3.14}). Though less elaborate than the proof presented in Section III, this alternative derivation (which generalizes the approach followed in paper II) permits a better understanding of why the multipole moments are given by (\ref{eq:3.12}).
 
We denote by $\Delta$ the {\it difference} between the field $h$, solution of the field equations (\ref{eq:2.14})-(\ref{eq:2.16}), and the finite part of the retarded integral of ${\cal M}(\Lambda)$ as given by the first term in (\ref{eq:3.11})~:

\begin{equation}
\Delta \equiv h - {\rm FP}_{B=0}\, \Box^{-1}_R [ \tilde r^B
{\cal M}(\Lambda)] \ . \label{eq:A1} 
\end{equation} 
Since $h$ is given by (\ref{eq:2.18}) this difference reads

\begin{equation}
\Delta = {16\pi G\over c^4} \Box ^{-1}_{R} \tau - {\rm FP}_{B=0}\, 
  \Box^{-1}_R [ \tilde r^B {\cal M}(\Lambda)] \ . \label{eq:A2} 
\end{equation} 
In the second term the finite part at $B=0$ is necessary because the multipole expansion ${\cal M}(\Lambda)$ is singular at the origin $r=0$. On the other hand $\tau$ in the first term of (\ref{eq:A2}) is regular all over $I\!\!R^4$, and therefore one can conveniently add into this term the finite part at $B=0$ without changing its numerical value (for convergent integrals the finite part simply gives back the value of the integral). Thus $\Delta$ can be equivalently re-written into the more useful form

\begin{equation}
\Delta = {\rm FP}_{B=0}\, \Box^{-1}_R \left[ \tilde r^B \left({16\pi G\over c^4} \tau - {\cal M}(\Lambda)\right) \right] \ . \label{eq:A3} 
\end{equation} 
Under this form $\Delta$ appears to be the (finite part of a) retarded integral of a source with spatially {\it compact} support. This readily follows from our assumption that $\tau={\cal M}(\tau)$ when $r>{\cal R}$ [indeed, apply $\Box$ on (\ref{eq:2.21})], and the fact that ${\cal M} (\Lambda)=16\pi G/c^4 {\cal M}(\tau)$ because $T$ has a compact support. Therefore the multipole expansion of $\Delta$ in the region $r>{\cal R}$ can be obtained directly from the standard formula valid for sources with compact support. This yields immediately

\begin{equation}
{\cal M}(\Delta) = - {4G\over c^4} \sum^{+\infty}_{l=0}
 {(-)^l\over l!} \partial_L \left\{ {1\over r} {\cal
 H}_L (t-r/c) \right\} \ ,   \label{eq:A4}
\end{equation} 
where the multipole moments are given by

\begin{equation}
 {\cal H}_L = {\rm FP}_{B=0}  \int d^3 {\bf y} |\tilde {\bf y}|^B y_L \,
\left(\tau - {c^4 \over 16\pi G}{\cal M}(\Lambda)\right) \ .   \label{eq:A5}
\end{equation} 
Now in the case of a slowly-moving source the zone of validity of the post-Newtonian expansion (or near-zone) covers the compact support of the source~: $d < {\cal R} \ll \lambda$ in the notation of Section II. Therefore both $\tau$ and ${\cal M}(\Lambda)$ in (\ref{eq:A5}) can be replaced by their post-Newtonian expansions, i.e.
        
\begin{equation}
 {\cal H}_L = {\rm FP}_{B=0}  \int d^3 {\bf y} |\tilde{\bf y}|^B y_L \,
\left(\overline{\tau} - {c^4 \over 16\pi G}\overline{{\cal M}(\Lambda)}\right)\ ,   \label{eq:A6}
\end{equation} 
Finally thanks to the structure of the post-Newtonian (or near-zone) expansion $\overline{{\cal M}(\Lambda)}$ as given by the right side of (\ref{eq:2.23}), we see that after integration over the angles the second term in (\ref{eq:A6}) is a sum of terms of the type ${\rm FP}_{B=0} B\int_0^{+\infty} d|{\bf y}| |\tilde{\bf y}|^B |{\bf y}|^{2+l+a} (\ln |\tilde{\bf y}|)^p$ multiplied by a function of time. All the latter radial integrals are zero by analytic continuation in $B$ [see the discussion after (\ref{eq:3.5})], thus we conclude that the second term in (\ref{eq:A6}) vanishes identically, and we recover the same result as obtained in (\ref{eq:3.12}):

\begin{equation}
 {\cal H}_L  = {\rm FP}_{B=0} \int d^3 {\bf y} |\tilde {\bf y}|^B y_L \overline{\tau}\ .   \label{eq:A7}
\end{equation} 

\section{The harmonicity algorithm}
\label{sec:apb}

The role of the tensor $v^{\mu\nu}$ defined by (\ref{eq:4.7}) is to
cancel out the divergence of the tensor $u^{\mu\nu}$ given by
(\ref{eq:4.1}), i.e.  $\partial_\nu (u^{\mu\nu} + v^{\mu\nu}) = 0$,
while at the same time being a solution of the source-free wave
equation, i.e.  $\Box v^{\mu\nu}=0$.  We first show the agreement
between (\ref{eq:4.7}) and the expression (2.12) in \cite{B97quad} which
is itself a slight modification of the earlier definition (4.13) in
paper I.  The divergence $\partial_\nu u^{\mu\nu}$ is given by
(\ref{eq:4.6}) in terms of the moments ${\cal G}^\mu_L$ which are
themselves decomposed in (\ref{eq:5.3}) into STF tensors $P_L$,
$^{(+)}Q_{L+1}$, $^{(0)}Q_L$ and $^{(-)}Q_{L-1}$.  Posing

\begin{mathletters} \label{eq:B1} \begin{eqnarray}
 A_L &=& {4G\over c^4}\, {(-)^l \over l !} P_L\ , \label{eq:B1a}  \\
 B_L &=& {4G\over c^4}\, {(-)^l\over l !}
     {-1\over l+1} {}^{(-)}Q_L\ , \label{eq:B1b} \\
 C_L &=& {4G\over  c^4}\, {(-)^l\over l !}
  \left[ -l {}^{(+)}Q_L + {l\over c^2(l+1)(2l+1)}
  {}^{(-)}\ddot Q_L \right]\ , \label{eq:B1c}\\
 D_L &=& {4G\over c^4}\,  {(-)^l\over l !}
  {}^{(0)} Q_L \ , \label{eq:B1d} \end{eqnarray}
\end{mathletters} 
we can re-write (\ref{eq:4.6}) as
 
\begin{mathletters} \label{eq:B2}
\begin{eqnarray}
  \partial_\nu u^{0\nu} &=&\sum_{l\geq 0} \partial_L \left({1\over
  r} A_L\right)\ , \label{eq:B2a}\\ \partial_\nu u^{i\nu}
  &=&\sum_{l\geq 0} \partial_{iL} \left({1\over r} B_L\right)
  \nonumber \\ 
&&+ \sum_{l\geq 1} \left\{ \partial_{L-1} \left({1\over r}
   C_{iL-1}\right) + \varepsilon_{iab} \partial_{aL-1} \left({1\over r}
   D_{bL-1}\right) \right\}\ .  \label{eq:B2b}
\end{eqnarray}
\end{mathletters}
On the other hand the tensor $v^{\mu\nu}$ given by (4.7) is easily transformed into

\begin{mathletters}
\label{eq:B3}
\begin{eqnarray}
 v^{00} &=& - {c\over r} \hbox{$\int$} A + \partial_a \left( {1\over r}
 \left[-c\hbox{$\int$} A_a+ c^2\hbox{$\int\!\!\int$} C_a -3B_a\right]
      \right)\ ,\label{eq:B3a}\\
 v^{0i} &=& {1\over r}\left[ -c \hbox{$\int$} C_i +{3\over c}\dot B_i\right]
  - c\varepsilon_{iab} \partial_a
 \left( {1\over r} \hbox{$\int$} D_b \right) -\sum_{l\geq 2}\partial_{L-1}
 \left( {1\over r} A_{iL-1} \right) \ , \label{eq:B3b} \\
 v^{ij} &=& - \delta_{ij} {1\over r} B + \sum_{l\geq 2} \left\{
   2 \delta_{ij}\partial_{L-1} \left( {1\over r} B_{L-1}\right) -6
   \partial_{L-2(i} \left( {1\over r} B_{j)L-2}\right) \right. \nonumber \\
 && +\partial_{L-2} \left( {1\over r} \left[ {1\over c}\dot A_{ijL-2}
  + {3\over c^2} \ddot B_{ijL-2} - C_{ijL-2} \right] \right)
   \nonumber \\
 && \left. - 2 \partial_{aL-2}\left( {1\over r}\varepsilon_{ab(i} D_{j)bL-2}
   \right) \right\}\ , \label{eq:B3c}
\end{eqnarray}
whose spatial trace is simply monopolar:

\begin{equation}
   v^{ii} = - {3\over r} B \ .\label{eq:B3d}
\end{equation}
\end{mathletters}
We thus have agreement with the definition proposed in the equations 
(2.11)-(2.12) in \cite{B97quad}.

Next we compare $v^{\mu\nu}$ with the earlier definition proposed 
in paper I, that we denote here by $q^{\mu\nu}$. Using the same notation 
as for (\ref{eq:B3}) we have

\begin{mathletters} \label{eq:B4}
\begin{eqnarray}
 q^{00} &=& - {c\over r} \hbox{$\int$} A + \partial_a \left( {1\over r}
 \left[ -c\hbox{$\int$} A_a+ c^2 \hbox{$\int\!\!\int$} C_a
 \right]\right)\ , \label{eq:B4a}\\
 q^{0i} &=& - {c\over r} \hbox{$\int$} C_i - c\varepsilon_{iab}
 \partial_a \left( {1\over r} \hbox{$\int$} D_b \right)-\sum_{l\geq
 2} \partial_{L-1} \left( {1\over r} A_{iL-1} \right) \ ,
 \label{eq:B4b} \\ 
q^{ij} &=& - \delta_{ij} \left[ {1\over r} B + \partial_a \left(
  {1\over r} B_a\right) \right] \nonumber \\ 
&& + \sum_{l\geq 2} \left\{ 2\delta_{ij} \partial_L \left( {1\over r}
    B_L\right) - 6\partial_{L-1(i} \left( {1\over r} B_{j)L-1} \right)
   \right. \nonumber \\ 
&& +\partial_{L-2} \left( {1\over r} \left[ {1\over c}\dot A_{ijL-2}
  + {3\over c^2} \ddot B_{ijL-2} - C_{ijL-2} \right] \right)
   \nonumber \\ && \left.- 2 \partial_{aL-2} \left( {1\over r}
 \varepsilon_{ab(i} D_{j)bL-2} \right) \right\}\ , \label{eq:B4c} 
\end{eqnarray} 
with spatial trace

\begin{equation}
 q^{ii} = - 3 \left[ {1\over r} B + \partial_a \left( {1\over r} B_a
     \right) \right] \ . \label{eq:B4d} 
\end{equation}
\end{mathletters} 
Subtracting
(\ref{eq:B4}) from (\ref{eq:B3}) we obtain 

\begin{mathletters}
\label{eq:B5} 
\begin{eqnarray}
  v^{00} - q^{00} &=& - 3 \partial_a \left( {1\over r} B_a\right)\ ,
    \label{eq:B5a}\\ 
v^{0i} - q^{0i} &=& {3\over rc} \dot B_i\ , \label{eq:B5b}\\ 
v^{ij} - q^{ij} &=& 3 \delta_{ij}\partial_a \left( {1\over r} B_a\right)
 - 6 \partial_{(i} \left( {1\over r} B_{j)} \right)\ , \label{eq:B5c} 
\end{eqnarray} 
\end{mathletters} 
which can be re-expressed in the form of the gauge transformation 

\begin{equation}
 v^{\mu\nu} - q^{\mu\nu} = \partial^\mu \varepsilon^\nu + \partial^\nu
 \varepsilon^\mu - \eta^{\mu\nu} \partial_\lambda \varepsilon^\lambda
 \ , \label{eq:B6} 
\end{equation} 
associated with the vector

\begin{mathletters} \label{eq:B7} 
\begin{eqnarray}
 \varepsilon^0 &=& 0 \ , \label{eq:B7a}\\ 
\varepsilon^i &=&  - {3\over r} B_i \ . \label{eq:B7b} 
\end{eqnarray} 
 \end{mathletters} 
Thus, had we used in (\ref{eq:4.8}) the tensor $q^{\mu\nu}$ instead of the tensor $v^{\mu\nu}$, i.e. had we considered instead of $Gh^{\mu\nu}_{\rm part.1}$ the different linearized metric

\begin{equation}
 Gh'^{\mu\nu}_{\rm part.1} = Gh^{\mu\nu}_{\rm part.1}
  + \partial^\mu \varepsilon^\nu +
   \partial^\nu \varepsilon^\mu - \eta^{\mu\nu}\partial_\lambda
  \varepsilon^\lambda \ , \label{eq:B8}
\end{equation} 
we would have obtained a dipole moment $Y_i'$ differing from the dipole
$Y_i$ as given by (\ref{eq:5.19}) [or (\ref{eq:5.8c})] with $l =1$ by
the formula

\begin{equation}
 Y_i' = Y_i - {3c^4\over 4G} B_i = Y_i - {3\over 2}{} ^{(-)}Q_i \ , 
  \label{eq:B10} 
\end{equation} 
with no other modifications whatsoever.  Thus the expression of the
dipole moment $Y_i'$ would not be ``uniform'' with the expressions of the other moments $Y_L$ for arbitrary $l\geq 2$. In this paper we have opted for the definition $v^{\mu\nu}$ instead of $q^{\mu\nu}$ for this reason, and also because the spatial trace $v^{ii}$ is simpler than the corresponding $q^{ii}$ [compare (\ref{eq:B3d}) and (\ref{eq:B4d})].

\end{document}